\documentclass[twocolumn]{autart}

\usepackage[utf8]{inputenc} 
 \usepackage[T1]{fontenc}    
 \usepackage{amsfonts}       
 \usepackage{amsmath} 
 \usepackage{amssymb}  
 \usepackage{mathrsfs}
 \usepackage{algorithm}
 \usepackage{algpseudocode}
 \algnewcommand{\Inputs}[1]{%
  \State \textbf{Inputs:}
  \Statex \hspace*{\algorithmicindent}\parbox[t]{.8\linewidth}{\raggedright #1}
}
\algnewcommand{\Initialize}[1]{%
  \State \textbf{Initialize:}
  \Statex \hspace*{\algorithmicindent}\parbox[t]{.8\linewidth}{\raggedright #1}
}

\usepackage[a4paper]{geometry}
 \usepackage{bm}
 \usepackage{array}
 \usepackage{stfloats}
 \usepackage{graphicx}
 \usepackage[colorlinks=true, allcolors=black]{hyperref}
 \DeclareMathAlphabet{\mathcal}{OMS}{cmsy}{m}{n}
 \DeclareSymbolFont{largesymbols}{OMX}{cmex}{m}{n}
 \usepackage[normalem]{ulem}
 
\usepackage[normalem]{ulem}

\newcommand{\mS}{\mathcal{S}}
\newcommand{\mX}{\mathcal{X}}
\newcommand{\mU}{\mathcal{U}}
\newcommand{\mR}{\mathcal{R}}
\newcommand{\R}{\mathbb{R}}
\newcommand{\mbE}{\mathbb{E}}
\newcommand{\mbP}{\mathbb{P}}

\newcommand{\mO}{\mathcal{O}}

\newcommand{\tx}{\tilde{x}}
\newcommand{\tX}{\tilde{X}}
\newcommand{\ty}{\tilde{y}}

\newcommand{\tw}{\tilde{w}}

\usepackage{pdfpages}
\usepackage{stmaryrd}

\newcommand{\OF}{\mathsf{F}}

\newcommand{\salf}[1]{\Phi_{n,\lambda}(#1)}

\newcommand{\defeq}{:=}

\newcommand{\innerp}[1]{\langle #1 \rangle}
\newcommand{\expect}[1]{\mathbb{E}\left( #1 \right)}
\newcommand{\expectw}[2]{\mathbb{E}_{#1}\left( #2 \right)}
\newcommand{\proba}[1]{\mathbb{P}\left( #1 \right)}

\newtheorem{definition}{Definition}[section]
\newtheorem{lemma}{Lemma}[section]
\newtheorem{assumption}{Assumption}
\newtheorem{remark}{Remark}[section]
\newtheorem{proposition}{Proposition}

\newtheorem{problem}{Problem}

\begin{document}

\begin{frontmatter}

\title{Probabilistic Reachability of Discrete-Time Nonlinear Stochastic Systems \thanksref{footnoteinfo}}
\thanks[footnoteinfo]{This paper was not presented at any IFAC 
meeting.}

\author[Atlanta]{Zishun Liu}\ead{zliu910@gatech.edu}, 
\author[CO]{Saber Jafarpour}\ead{Saber.Jafarpour@colorado.edu}, 
\author[Atlanta]{Yongxin Chen}\ead{yongchen@gatech.edu},              

\address[Atlanta]{Georgia Institute of Technology, Atlanta, GA 30332, USA}      
\address[CO]{University of Colorado Boulder, Boulder, CO 80309, USA}


\begin{abstract}
  In this paper we study the reachability problem for discrete-time nonlinear stochastic systems. 
  Our goal is to present a unified framework for calculating the probabilistic reachable set of discrete-time systems in the presence of both deterministic input and stochastic noise. 
  By adopting a suitable separation strategy, the probabilistic reachable set is decoupled into a deterministic reachable set and the effect of the stochastic noise. 
  To capture the effect of the stochastic noise, in particular sub-Gaussian noise, we provide a probabilistic bound on the distance between a stochastic trajectory and its deterministic counterpart. 
  The key to our approach is a novel energy function called the Averaged Moment Generating Function, which we leverage to provide a high probability bound on this distance.
  We show that this probabilistic bound is tight for a large class of discrete-time nonlinear stochastic systems and is exact for linear stochastic dynamics. 
  By combining this tight probabilistic bound with the existing methods for deterministic reachability analysis, we propose a flexible framework that can efficiently compute probabilistic reachable sets of stochastic systems. 
  We also provide two case studies for applying our framework to Lipschitz bound reachability and interval-based reachability. Three numerical experiments are conducted to validate the theoretical results.
\end{abstract}
\end{frontmatter}

\section{Introduction}
\label{sec:introduction}
Reachability analysis is a classical problem that studies how a dynamical system evolves over time under uncertainties in its initial conditions and inputs. 
It naturally appears in diverse applications, including model checking, safety verification of autonomous systems, and controller synthesis. 
It is known that obtaining the exact reachable sets of general dynamical systems requires computing an infinite number of their trajectories, which is computationally intractable~\cite{XC-SS:22}. 
Consequently, numerous frameworks have been developed in the literature to efficiently approximate the reachable sets for both continuous-time and discrete-time dynamical systems. 
While continuous-time models are commonly used to represent physical processes, discrete-time models are better suited for systems involving digital implementation and real-time computation~\cite{SVR-ECK-DQM-JL:06,meng2021reactive,cosner2024bounding}.
In this paper, we focus on the reachability analysis of discrete-time systems.


%

For deterministic systems, reachability studies typically handle the system in the presence of bounded input and provide worst-case approximation for the reachable set \cite{prajna2007framework}. Approaches in this line can be roughly divided into three categories. 
Dynamical programming approaches use a game-theoretic perspective to compute reachable sets of discrete-time systems~\cite{AA-MP-JL-SS:08,SS-JL:10}. 
However, the computational cost of these approaches makes them intractable for reachability analysis of large-scale systems.  
Set propagation methods rely on propagating a specific family of sets to over-approximate the reachable sets of the system (such as polytopes, ellipsoids, and zonotopes)~\cite{SVR-ECK-DQM-JL:06,MK-DH-CNJ:14}.
Despite their computational efficiency, these approaches are either applicable to a special class of systems or lead to overly conservative estimates of reachable sets.
Recently Lipschitz bound~\cite{ZH-SM:12,koller2018learning} and interval analysis~\cite{LJ-MK-OD-EW:01,SC-MA:15b} have been used as flexible and computationally efficient set propagation methods for reachability analysis of large-scale deterministic systems.
%
Another category is simulation-based techniques, by using which the reachable set is estimated through numerical simulations \cite{murat2020dd,chuchu2017simulation,JM-MA:15,ZH-SM:12,fan2016C2E2}. 
All these methods constitute a toolset for reachability analysis on deterministic systems with bounded input. 

Many real-world applications exhibit uncertainties that are highly variable and unpredictable. To analyze the propagation of these uncertainties through the system, it is more effective to describe them using stochastic disturbances.
 When the stochastic disturbance is bounded, the deterministic reachability methods based on the worst-case analysis can be used to estimate reachable sets of the system. However, this approach can lead to trivial or overly-conservative estimates of reachable sets as it offer excessive robustness to the worst-case adversarial noise, which rarely happens in practice \cite{cosner2024bounding}.
 When the stochastic disturbance is unbounded, such as Gaussian noise, the set of all possible stochastic trajectories can be unbounded.
%
To better capture the effects of stochastic disturbances, reachability analysis in stochastic systems focuses on the \textit{probabilistic reachable set}, which refers to the set that any possible trajectory
starting from an initial set can reach with high probability (e.g., 99.9\%).


The probabilistic reachability of discrete-time systems has been studied using various techniques. 
A prevalent approach is to approximate the stochastic disturbance with a bounded input, for instance, by establishing a high-probability bound on the stochastic noise~\cite{10068731,lew2021sampling}. This converts the probabilistic reachability analysis into a worst-case deterministic reachability problem.
%
However, this strategy can lead to a conservative estimates of the probabilistic reachable sets for high probability levels over long time periods (see discussions in Section \ref{subsec: comparison}).
%
Optimization-based approaches leverage methods such as dynamic programming to accurately compute the probabilistic reachable set~\cite{AA-MP-JL-SS:08,SS-JL:10,APV-MKO:21}. Despite the high accuracy of these approaches, they can become computationally inefficient for complex and large-scale systems. 
Another relevant line of literature for reachability analysis is the use of Lyapunov functions~\cite{hewing2018stochastic,black2024risk} or control barrier function~\cite{santoyo2021barrier,chern2021safe} for measuring the probability of a trajectory staying in a level set. However, these functional approaches are only efficient in restrictive scenarios. For general nonlinear systems, finding a probabilistic reachable set that can be verified through these approaches can be difficult. 
%
%
 
%


In this paper, we develop a framework for probabilistic reachability analysis of discrete-time stochastic systems under both bounded input and stochastic disturbances.
Inspired by~\cite{szy2024TAC}, we establish a separation strategy that decouples the effects of deterministic inputs and stochastic uncertainties on the probabilistic reachable set of systems.  
The effect of deterministic input is captured using reachable sets of an associated deterministic system and the effect of stochastic uncertainty is represented using the distance between trajectories of the stochastic system and the associated deterministic system termed as stochastic deviation. 
By leveraging an energy function called the Averaged Moment Generating Function (AMGF), we prove a tight probabilistic bound (Theorem \ref{thm: high-prob}) on the stochastic deviation for general nonlinear discrete-time systems. 
This bound 
is sharper than bounds achieved by the worst-case analysis methods such as conformal prediction. Moreover, our bound coincides with the classical bound on stochastic deviation for linear stochastic systems under the same assumptions.
Our bound is applicable to systems subject to sub-Gaussian stochastic noise, which includes a wide range of typical noise such as Gaussian, truncated Gaussian and uniform noise. 
We show that this bound cannot be improved further without additional assumptions. 

Our framework offers tremendous flexibility in finding probabilistic reachable sets of stochastic systems, as it can incorporate any deterministic reachability methods for approximating the effect of deterministic inputs with high probability tight bounds on the stochastic deviation. 
%
 Due to the separation strategy and the tightness of the stochastic deviation, the computational efficiency and tightness of the probabilistic reachable set only rely on the over-approximation of the deterministic reachable set. Consequently, analyzing the reachability of the associated deterministic system is all we need to obtain a good probabilistic reachable set, and computational heavy work on the stochastic system can be avoided. In particular, we combine our framework with two computationally efficient deterministic reachability approaches, namely Lipschitz bound reachability and interval-based reachability, and provide explicit expressions for the probabilistic reachable sets of discrete-time stochastic systems.

Finally, the contribution of this probabilistic bound goes beyond reachability analysis. Our tight probabilistic bound of stochastic deviation is the first non-conservative result that can quantitatively describe the behavior of a general nonlinear discrete-time system subject to sub-Gaussian noise. This tight bound is poised to have profound implications in the broader range of research and applications such as safety-critical control, finance, statistics, machine learning, etc.

The rest of the paper is organized as follows. We provide background knowledge about deterministic reachability, formulate the probabilistic reachability problem, and presents our overall strategy in Section \ref{sec: bg}. In Section \ref{sec: expectation} we provide the expectation bound on the stochastic deviation introduced in Section \ref{sec: bg} and points out its limitations. To overcome the limitations, we introduce the main technical contribution termed Averaged Moment Generating Function in Section \ref{sec: high-prob bound}, and make use of it to prove the probabilistic bound on the stochastic deviation. We also show the tightness of our result and compare it with the performance of the worst-case analysis in this section. 
Based on the theoretical result of Section \ref{sec: high-prob bound}, Section \ref{sec: PRS} provides the result of the probabilistic reachable set and gives two related case studies. Numerical experiments are displayed in Section \ref{sec: simulations}. Section \ref{sec: conclusion} concludes the paper.

{\bf Notations.} For a vector $x\in \R^n$, $\|x\|$ denotes its Euclidean norm ($l_2$ norm) and $\innerp{x,y}$ denotes the corresponding inner product. For a matrix $A$, $\|A\|$ denotes its induced norm. Given two sets $A,B$, we define the Minkowski sum of the sets $A$ and $B$ by $A\oplus B = \{x+y: x\in A,~ y\in B\}$. Throughout the paper, we use $\mbE$ to denote expectation, $\mbP$ to denote probability, $\mathcal{B}^n(r,y)$ to denote the ball $\{x\in\R^n: \|x-y\|\leq r\}$, and $\mS^{n-1}$ to denote the unit sphere of $\R^n$. For a random variable $X$, $X\sim G$ means $X$ is independently drawn from the distribution $G$ and $X\sim \mS^{n-1}$ means $X$ is drawn uniformly from $\mS^{n-1}$. $N(\mu,\Sigma)$ denotes the Gaussian distribution with mean $\mu$ and covariance matrix $\Sigma$. Given a function $f:\R\to \R$ and a non-negative function $g:\R\to \R_{\ge 0}$, we write $f(x)=\mathcal{O}(g(x))$, if there exist constants $M>0$ and $x_0\in \R$ such that $|f(x)|\le Mg(x)$, for every $x\ge x_0$. 

\section{Problem Statement and Preliminaries}\label{sec: bg}
Consider the discrete-time stochastic system 
\begin{equation}\label{sys: d-t ss}
     X_{t+1}=f(X_t,u_t,t)+w_t
\end{equation}
where $X_t\in \R^n$ is the state, $u_t\in \R^{p}$ is the input, $w_t\in\R^n$ is the stochastic disturbance, and $f: \R^n\times\R^p\times \R_+\to\R^n$ is a parameterized vector field. The input $u_t$ can be either control input or deterministic disturbance depending on the application. 
In this paper, we impose the following standard Lipschitz condition on $f$~\cite{novara2013direct}.
\begin{assumption}\label{ass: Lipschitz f}
    At every time $t\geq0$, there exists $L_t\geq0$ such that, for every $x,y\in \R^n$ and every $u\in \R^p$, 
        $$\|f(x,u,t) - f(y,u,t)\| \leq L_t\|x-y\|.$$
\end{assumption}
%
This paper aims to establish an effective method of reachability analysis for the stochastic system \eqref{sys: d-t ss}. In this section, we formulate our probabilistic reachability problem on the stochastic system \eqref{sys: d-t ss} following a brief review of deterministic reachability. We then present our overall strategy for addressing the problem.

\subsection{Deterministic Reachable Set}\label{subsec: DRS}
Computing the reachable sets for deterministic systems 
is a fundamental problem in dynamical systems and control theory. Consider the deterministic system 
\begin{equation}\label{sys: d-t ds}
    x_{t+1}=f(x_t,u_t,t),
\end{equation}
which can be viewed as a noiseless version of the stochastic system \eqref{sys: d-t ss}. 
The deterministic reachable set (DRS) at time $t$ is the set in the state space that the system \eqref{sys: d-t ds} can reach, starting from an initial configuration, under all possible inputs within a specified time.
\begin{definition}[DRS]\label{def: DRS}
 Consider the system~\eqref{sys: d-t ds} with initial set $\mathcal{X}_0\subseteq \R^n$ and input set $\mathcal{U}\subseteq \R^p$. The \textit{deterministic reachable set} of~\eqref{sys: d-t ds} at time $t$ starting from $\mathcal{X}_0$ with inputs in $\mathcal{U}$ is
    \begin{equation}\label{eq: DRS}
    \mathcal{R}_t = \left\{ x_t \middle|
    \begin{aligned}
    &\tau\mapsto x_\tau \mbox{ is a trajectory of~\eqref{sys: d-t ds}}\\ &\mbox{ with } x_0\in \mathcal{X}_0 \mbox{ and } u_\tau: \mathbb{Z}_{\ge 0}\to \mathcal{U}
    \end{aligned}
    \right\}
\end{equation}
\end{definition}

In general, computing the exact DRS of a dynamic system is challenging \cite{CM:90}. Therefore, most methods in reachability analysis focus on providing over-approximation of DRS~\cite{XC-SS:22}. A set $\overline{\mathcal{R}}_t\subseteq \R^n$ is an over-approximation of the DRS $\mR_t$ if
\begin{align*}
    \mathcal{R}_t\subseteq \overline{\mathcal{R}}_t.
\end{align*}
The calculation of $\overline{\mR}_t$ depends on the properties of system dynamics $f$ \cite{FB-CTDS}, and types of input $u_t$ \cite{prajna2007framework}. In Section \ref{sec: PRS}, we review two approaches to compute $\overline{\mR}_t$: Lipschitz bound reachability and interval-based reachability.

\subsection{Probabilistic Reachable Set}
\begin{figure}
\centering
\includegraphics[width =0.6\linewidth]{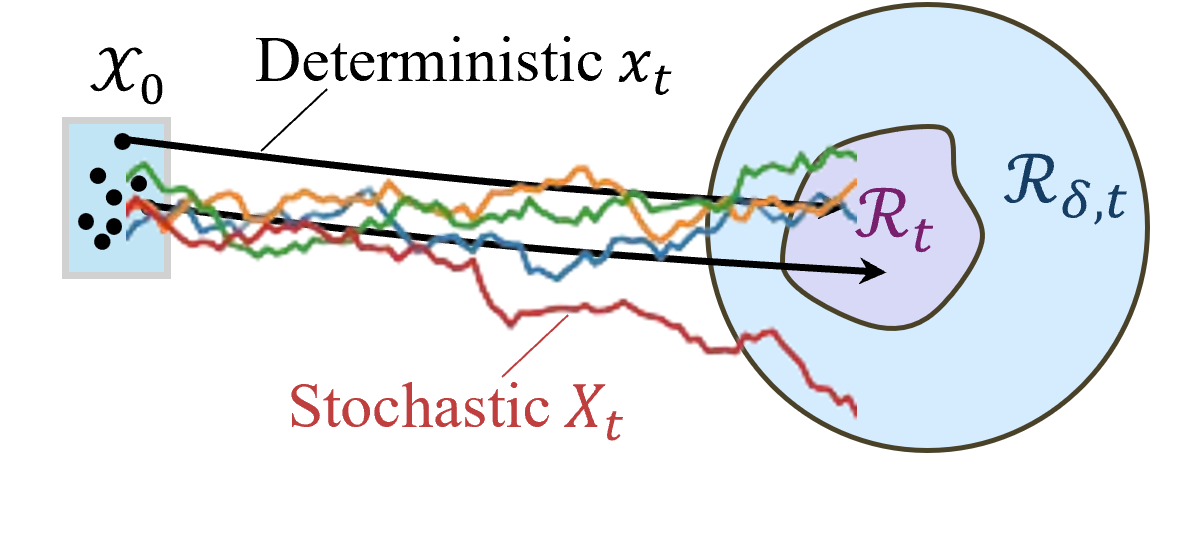}
\caption{An illustration of $\delta$-PRS at time $t$. Here $\mR_{\delta,t}$ is the $\delta$-PRS of the stochastic system \eqref{sys: d-t ss}, whose trajectories are in color. $\mR_t$ is the DRS of the associated deterministic system \eqref{sys: d-t ds}, whose trajectories are in black.}
\label{fig: prs}
\end{figure}
We are interested in characterizing the reachable set of \eqref{sys: d-t ss}. We focus on the setting where the disturbance $w_t$ is \textit{sub-Gaussian}. The sub-Gaussian distribution is quite general and includes a wide range of distributions such as Gaussian, uniform, and any zero-mean distributions with bounded support.
\begin{definition}[sub-Gaussian] \label{def: subG}
    A random variable $X\in\R^n$ is said to be sub-Gaussian with variance proxy $\sigma^2$, denoted as $X\sim subG(\sigma^2)$, if $\mbE(X)=0$ and $X$ satisfies
\begin{equation}\label{eq: subG}
    \mbE_X\left(e^{\lambda \innerp{\ell,X}}\right)\leq e^{\frac{\lambda^2\sigma^2}{2}},~\forall \lambda\in\R, ~\forall \ell\in\mS^{n-1}.
\end{equation}
\end{definition}
\begin{assumption}\label{ass: bounded sigma}
    The disturbance $w_t\sim subG(\sigma_t^2)$ with some finite $\sigma_t\geq0$ for every $t\geq0$.
\end{assumption}

The theory for deterministic reachability analysis is not effective in handling stochastic disturbances. 
When $w_t$ is unbounded sub-Gaussian noise, the DRS in the sense of \eqref{eq: DRS} is often trivial. To see this, consider the system $X_{t+1}=X_t+w_t$ where $w_t$ is a zero-mean Gaussian noise. The state $X_t$ is a Gaussian vector whose possible range is the entire $\R^n$ space and thus $\mathcal{R}_t=\R^n$. When $w_t$ is bounded, the deterministic reachability analysis ignores the statistics of the disturbance and treats it in a worst-case manner, leading to conservative results. In reality, the worst-case realization of stochastic disturbance occurs with extremely low probability and is thus not representative. 
For these reasons, we turn to the concept of \textit{probabilistic reachable set}, which refers to the set that any possible trajectory starting from the initial state set $x_0\in\mX_0$ and with the input $u_\tau\in\mU$ can reach with high probability.
\begin{definition}[$\delta$-PRS]\label{def: PRS}
Consider the stochastic system \eqref{sys: d-t ss} with the initial set $\mX_0\subseteq\R^n$ and the bounded input set $\mU\subseteq\R^p$. Given $\delta\in (0,1]$ and $t\in\mathbb{Z}_{\ge 0}$, the set $\mathcal{R}_{\delta,t}\subseteq \R^n$ is said to be a $\delta$-probabilistic reachable set ($\delta$-PRS) of the system \eqref{sys: d-t ss} at time $t$, if for any $x_0\in\mX_0$ and $u_t: \mathbb{Z}_{\ge 0}\to\mU$, it holds that
    \begin{equation}\label{eq: def p-PRS}
        \proba{X_t\in\mR_{\delta,t}}\geq 1-\delta.
    \end{equation}
\end{definition}
Apparently, as can be seen from the definition, $\delta$-PRS is not unique.
Indeed, if $\mR_{\delta,t}$ is a $\delta$-PRS, then any $\mR_{\delta,t}'\supseteq\mR_{\delta,t}$ is also a $\delta$-PRS. We say $\mR_{\delta,t}$ is a \textit{tighter} $\delta$-PRS than $\mR_{\delta,t}'$ if $\mR_{\delta,t}\subseteq \mR_{\delta,t}'$. An illustration of DRS and $\delta$-PRS is in Figure \ref{fig: prs}. 
In many applications like safety-critical control, it is desirable to have a tight $\delta$-PRS, and a loose $\delta$-PRS can result in very conservative strategies. Therefore,  the main goal in probabilistic reachability is to find a tight $\delta$-PRS. 
\begin{problem}\label{problem: prs}
    Find an as tight as possible $\delta$-PRS $\mR_{\delta,t}$ of the stochastic system \eqref{sys: d-t ss} under Assumptions \ref{ass: Lipschitz f} and \ref{ass: bounded sigma}.
\end{problem}


\begin{figure}
\centering
\includegraphics[width =0.9\linewidth]{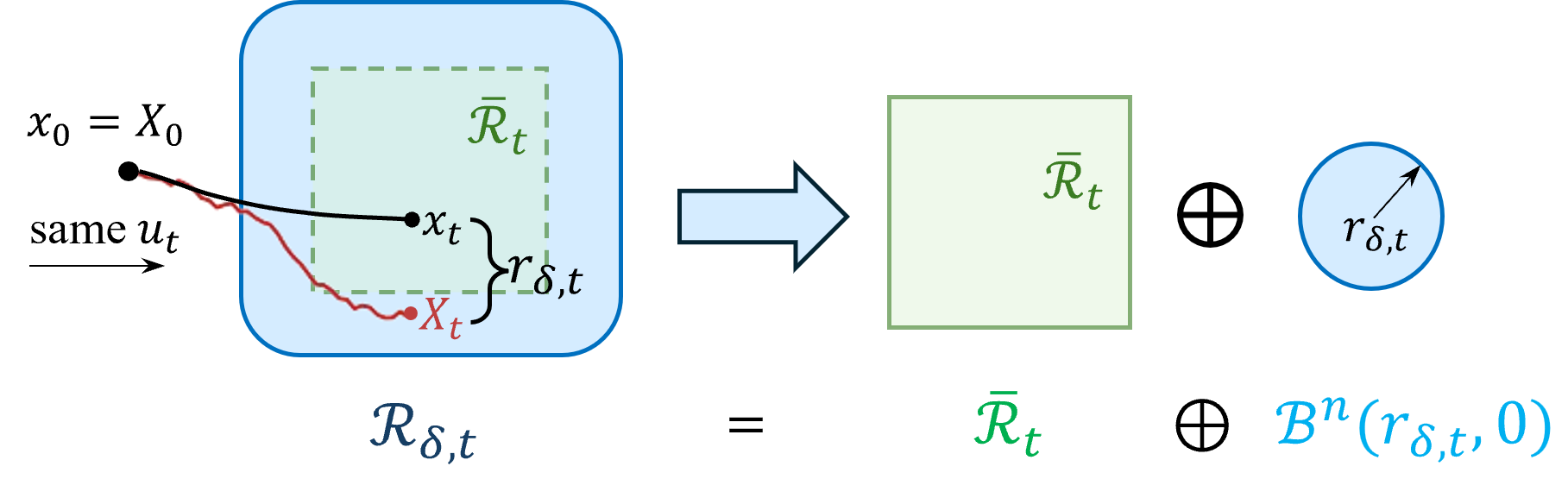}
\caption{An illustration of separation strategy. Here $\mR_{\delta,t}$ is the $\delta$-PRS of the stochastic system \eqref{sys: d-t ss}, whose trajectory is $X_t$ in red. $\overline{\mR}_t$ is the over-approximation of the DRS of the associated deterministic system \eqref{sys: d-t ds}, whose trajectory is $x_t$ in black. The Minkowski sum corresponds to Proposition \ref{prop: separation}.}
\label{fig: separation}
\end{figure}

\subsection{Separation Strategy and Main Challenge}

The trajectories of the stochastic system \eqref{sys: d-t ss} are driven by both deterministic terms ($f$,$u_t$) and stochastic disturbances ($w_t$). The impacts of these two parts on the trajectories are relatively independent and can be addressed separately. Building on this intuition, we proposed a strategy termed the \textit{separation strategy} in our recent work \cite{szy2024TAC} for probabilistic reachability analysis for continuous-time systems.

This separation strategy turns out to be extendable to the discrete-time setting. We say $X_t$ and $x_t$ are \textit{associated} trajectories if they have the same initial state $x_0$ and the same input $u_t$. The separation strategy states that the $\delta$-PRS can be separated into a deterministic part encoded by the DRS of \eqref{sys: d-t ds}, and a stochastic part, which is characterized as the distance $\|X_t-x_t\|$ between the associated trajectories, as formalized below. 
\begin{proposition}\label{prop: separation}
Consider the stochastic system \eqref{sys: d-t ss} with its associated deterministic system \eqref{sys: d-t ds}.
Given $\mathcal{X}_0\subseteq \R^n$, $\mathcal{U}\subseteq \R^p$, let $X_t$ and $x_t$ be the associated trajectories with inputs $u_t\in\mU$ and the initial state $x_0\in\mX_0$. If for any $t\geq0$, $x_0\in\mX_0$, $u_t\in\mU$ and $\delta>0$, $\exists r_{\delta,t}>0$ such that
   \begin{equation}\label{eq: lemma traj2set}
       \mathbb{P}\left(\|X_t-x_t\|\leq r_{\delta,t}\right)\geq 1-\delta,
   \end{equation}
   then for any over-approximation $\overline{\mR}_t$ of the DRS $\mR_t$ of \eqref{sys: d-t ds}, \begin{equation}\label{eq:separation}
        \overline{\mR}_t\oplus\mathcal{B}^n(r_{\delta,t},0)
    \end{equation}
   is a $\delta$-PRS of the system \eqref{sys: d-t ss}.
\end{proposition}
\begin{pf}
Let $X_t$ be any trajectory of \eqref{sys: d-t ss} associated with a trajectory $x_t$ of \eqref{sys: d-t ds}, then, by \eqref{eq: lemma traj2set}, $X_t\in \mathcal{B}^n(r_{\delta,t},x_t)$
with probability at least $1-\delta$. By the definition of $\overline{\mathcal{R}}_t$, $x_t\in\overline{\mathcal{R}}_t$. Therefore, by the definition of the Minkowski sum~\cite{weibel2007minkowski}, with probability at least $1-\delta$,
    \[
        X_t\in\overline{\mR}_t\oplus\mathcal{B}^n(r_{\delta,t},0),
    \]
which completes the proof.
\end{pf}
We term the difference $\|X_t-x_t\|$ between associated trajectories {\em stochastic deviation}. This separation strategy decomposes the probabilistic reachability analysis problem into two parts: approximate the DRS of \eqref{sys: d-t ds} and estimate the probabilistic bound $r_{\delta,t}$ of the stochastic deviation. Once a bound $r_{\delta,t}$ of the stochastic deviation is obtained, one can combine it with any existing deterministic reachability method to approximate the $\delta$-PRS. 

It is apparent that the size of the $\delta$-PRS $\overline{\mR}_t\oplus\mathcal{B}^n(r_{\delta,t},0)$ increases with $r_{\delta,t}$. Therefore, to ensure that $\overline{\mR}_t\oplus\mathcal{B}^n(r_{\delta,t},0)$ is not an overly-conservative $\delta$-PRS of \eqref{sys: d-t ss}, it is crucial to establish a probabilistic bound $r_{\delta,t}$ for the stochastic deviation that is as tight as possible. This is the main challenge addressed in this paper.

\begin{problem}[Stochastic Deviation]\label{problem: vib}
    Establish an as tight as possible probabilistic bound $r_{\delta,t}$ of the stochastic deviation $\|X_t-x_t\|$ associated with systems \eqref{sys: d-t ss}-\eqref{sys: d-t ds} under Assumptions \ref{ass: Lipschitz f} and~\ref{ass: bounded sigma}.  
\end{problem}

\section{Expectation Bound and Limitations}\label{sec: expectation}
%
Our first attempt for addressing Problem~\ref{problem: vib} is to adapt the expectation bound in our recent work \cite{szy2024TAC} to discrete-time systems. By applying the classical Markov inequality, the expectation bound translates into the probabilistic bound on stochastic deviation $\|X_t-x_t\|$. In this section, we present this approach and highlight its limitations.
%

\subsection{Expectation Bound on Stochastic Deviation}

We employ a similar technique to that in~\cite{szy2024TAC,QCP-NT-JJS:09} to analyze the stochastic deviation $\|X_t-x_t\|$ and establish bounds on $\mbE(\|X_t-x_t\|^2)$. The result is presented in the following proposition.
\begin{proposition}\label{prop: bound of E}
    Consider the stochastic system \eqref{sys: d-t ss} and its associated deterministic system \eqref{sys: d-t ds}. Suppose that Assumptions \ref{ass: Lipschitz f} and \ref{ass: bounded sigma} hold. Given $x_0\in \R^n$ and $u_t: \mathbb{Z}_{\ge 0}\to \R^p$ be an input sequence. Let $X_t$ (resp. $x_t$) be the trajectory of \eqref{sys: d-t ss} (resp. \eqref{sys: d-t ds}) with the input $u_t$ and the initial state $x_0$. Then,
    \begin{equation}\label{eq: thm E}
        \mbE\left(\|X_t-x_t\|^2\right)\leq n\Psi_t
    \end{equation}
     where
     \begin{equation}\label{eq: Psi val}
         \Psi_t=\psi_{t-1}\sum_{k=0}^{t-1}\sigma_{k}^2\psi_k^{-1},\quad 
             \psi_t=\prod_{k=0}^{t}L_k^{2}.
     \end{equation}
     
\end{proposition}

\begin{pf}
    Define $v_t=X_t-x_t$, $\beta_t=f(X_t,u_t,t)-f(x_t,u_t,t)$ and $V(\cdot)=\|\cdot\|^2$, then by \eqref{sys: d-t ss} and \eqref{sys: d-t ds} we have
\begin{equation}
     v_{t+1}=\beta_t + w_t.
\end{equation}
It follows that
\begin{equation}\label{eq: Vt+1=}
     V(v_{t+1})=V(\beta_t) + V(w_t) + 2\innerp{\beta_t,w_t}.
\end{equation}
Take the expectation of \eqref{eq: Vt+1=}. By Assumption \ref{ass: Lipschitz f}, $\expect{V(\beta_t)}\leq L_t^2 \expect{v_t}$. Since $w_t$ is independent from $\beta_t$, $\expect{\innerp{\beta_t,w_t}}=0$. Let $w_t=\begin{bmatrix} w_{t,1} & \cdots & w_{t,n}
\end{bmatrix}^{\top}$, then by Definition \ref{def: subG}, each entry $w_{t,i}$ of $w_t$ is sub-Gaussian with variance proxy $\sigma_t^2$.
Therefore,
\begin{equation} \label{eq: EV_t+1<=}
\begin{split}
     \mbE\left(V(v_{t+1})\right)&\leq L_t^2\mbE\left(V(v_t)\right)+\sum_{i=1}^n\expect{w_{t,i}^2}+0 \\
     &=L_t^2\mbE\left(V(v_t)\right)+\sum_{i=1}^n{\text{Var}(w_{t,i})} \\
     &\leq L_t^2\mbE\left(V(v_t)\right)+n\sigma_t^2, ~V(v_0)=0,
\end{split}
\end{equation}
where the last ``$\leq$'' uses the classical result that $\text{Var}(X)\leq \sigma^2$ if $X\in\mathbb{R}$ and $X\sim subG(\sigma^2)$ \cite{Alessandro2018lecture}. Solving the linear difference inequality \eqref{eq: EV_t+1<=} we obtain
\begin{equation}\label{eq: EV^2<=}
    \mbE\left(V(v_t)\right)\leq n\Psi_t,
\end{equation}
where $\Psi_t$ is defined in \eqref{eq: Psi val}. 
This completes the proof.
\end{pf}

\subsection{Limitation of The Expectation Bound}
The probabilistic bound directly derived from the expectation bound \eqref{eq: thm E} can be loose. By Markov's Inequality, for any $r>0$,
    \begin{equation}\label{eq: Markov}
    \begin{split}
        \proba{\|X_t-x_t\|\leq r} =& \proba{\|X_t-x_t\|^2\leq r^2} \\
        \geq &1-\frac{\expect{\|X_t-x_t\|^2}}{r^2}.
    \end{split}
    \end{equation}
Let $r=\sqrt{\frac{n}{\delta}\Psi_t}$ and apply Proposition \ref{prop: bound of E} to \eqref{eq: Markov}, then
\begin{equation}\label{eq: sq-root bound}
    \proba{\|X_t-x_t\|\leq \sqrt{\frac{n\Psi_t}{\delta}}}\geq 1-\delta.
\end{equation}

Given a probability level $1-\delta$, \eqref{eq: sq-root bound} provides a probabilistic bound on stochastic deviation that has $\mO(\sqrt{1/\delta})$ dependence on $\delta$. This is much worse than that for linear systems. To see this, consider a linear dynamics $X_{t+1}=AX_t+Bu_t+w_t$ which corresponds to \eqref{sys: d-t ss} with $f(x,u,t)=Ax+Bu$.
It satisfies Assumption \ref{ass: Lipschitz f} with $L_t\equiv \|A\|$, for every $t\ge 0$. For associated $X_t$ and $x_t$, by linearity, 
\begin{equation} \label{eq: lin Xt-xt}
\begin{split}
    &X_{t+1}-x_{t+1} = A(X_t-X_t) + w_t,\quad X_0-x_0=0 \\
    &\Rightarrow \,X_t-x_t = \sum_{k=0}^{t-1} A^kw_{t-1-k}.
\end{split}
\end{equation}
By the sum-up property of independent sub-Gaussian variables \cite{rigollet2023high}, \eqref{eq: lin Xt-xt} implies that  $X_t-x_t$ is sub-Gaussian with variance proxy 
\begin{equation}\label{eq: cov lin Xt-xt}
    \sigma^2(X_t) = \sum_{k=0}^{t-1} \|A\|^{2k}\sigma_k^2= \Psi_t.
\end{equation}

Sub-Gaussian random variables enjoy a strong norm concentration property stated below.
Note that the dependence $\mathcal{O}(\sqrt{n})$ and $\mathcal{O}(\sqrt{\log(1/\delta)})$ in Lemma \ref{lemma: concentration} are tight for sub-Gaussian distributions but the coefficients can be improved. We refer to \cite[Chapter 1.4]{rigollet2023high}\cite{szy2024TAC} for more details.
 \begin{lemma}[Norm Concentration]\label{lemma: concentration}
    Let $X\in\R^n$ be a sub-Gaussian random variable with variance proxy $\sigma^2$, then with any $\delta\in(0,1)$ and $\varepsilon\in(0,1)$, 
    \begin{equation}\label{eq: concentration norm}
        \|X\|\leq \sqrt{\sigma^2(\varepsilon_1n+\varepsilon_2\log(1/\delta))}
    \end{equation}
    holds with probability at least $1-\delta$, where 
    \begin{equation}\label{eq: epsilon val}
        \varepsilon_1=\frac{2\log(1+2/\varepsilon)}{(1-\varepsilon)^2},~ \varepsilon_2=\frac{2}{(1-\varepsilon)^2}.
    \end{equation}
\end{lemma}
 
 Plugging \eqref{eq: cov lin Xt-xt} into Lemma \ref{lemma: concentration}, we conclude that when the system \eqref{sys: d-t ss} is linear, the probabilistic bound on the stochastic deviation becomes 
\begin{equation} \label{eq: lin sv bound}
     r_{\delta,t}^{(lin)}= \sqrt{\Psi_t(\varepsilon_1n+\varepsilon_2\log(1/\delta))},
\end{equation}
which has an $\mO(\sqrt{\log(1/\delta))}$ dependence on $\delta$. When $\delta$ is sufficiently small (e.g., $\delta=10^{-6}$), which is crucial for safety-critical systems, $\sqrt{\log(1/\delta))}$ is significantly smaller than the term $\sqrt{1/\delta}$ in \eqref{eq: sq-root bound} (e.g., $2.6$ v.s. $10^3$). 

Thus, a significant gap between the result \eqref{eq: sq-root bound} for nonlinear dynamics and the probabilistic bound \eqref{eq: lin sv bound} for linear dynamics exists. This points to the question: is the gap fundamental or an artifact of the analysis?


\section{Probabilistic Bound on Stochastic Deviation}\label{sec: high-prob bound}
In this section, we answer the aforementioned question by establishing a probabilistic bound for the stochastic deviation $\|X_t-x_t\|$ of order $\mO(\sqrt{\log (1/\delta)})$ for general nonlinear stochastic systems \eqref{sys: d-t ss} under Assumptions \ref{ass: Lipschitz f} and~\ref{ass: bounded sigma}. We further show the tightness of our result and its advantages over the worst-case analysis for this problem.

\subsection{Motivation: Moment Generating Function}
The limitation of the expectation bound \eqref{eq: thm E} primarily lies in the quadratic Lyapunov function $V_t=\|X_t-x_t\|^2$. The analysis focuses only on the evolution of the second order moment $\mathbb{E}(\|X_t-x_t\|^2)$, 
so its upper bound at best guarantees a probabilistic bound for $\|X_t-x_t\|$ of order $\mO(\sqrt{1/\delta})$ via Markov inequality. In contrast, in Definition \ref{def: subG}, the function on the left-hand side of \eqref{eq: subG} captures arbitrarily high-order moments of a random variable, and its boundness leads to the norm concentration property of sub-Gaussian variables. This function is known as \textit{Moment Generating Function} (MGF) \cite[Chapter 1.1]{rigollet2023high}
\begin{equation}\label{eq: MGF}
    \mbE_X\left(M_{\lambda,\ell}(X)\right)\defeq \mbE_X\left(e^{\lambda\innerp{\ell,X}}\right),\quad \ell\in\mS^{n-1}.
\end{equation} 

Thus, to establish a tight probabilistic bound on the stochastic deviation, a potential approach is to upper bound the MGF of $X_t-x_t$ so that $X_t-x_t$ is sub-Gaussian and Lemma \ref{lemma: concentration} can be applied. 
Unfortunately, this is infeasible. For associated $X_t$ and $x_t$, $\mbE(X_t)\neq x_t$ when the systems are nonlinear, because
\begin{equation*}
    \begin{split}
        &\mbE(X_{t+1})= \mbE(f(X_t,u_t,t)) + \mbE(w_t) \\
        =&\mbE(f(X_t,u_t,t)) \neq f(\mbE(X_t),u_t,t),
    \end{split}
\end{equation*}
where the last ``$\neq$'' can be improved to ``$=$'' only if $f$ is linear. Therefore, the upper bound on $\mbE\left(M_{\lambda,\ell}(X_t-x_t)\right)$ is influenced by the value of $\expect{\innerp{\ell, X_t-x_t}}$ \cite[Chapter 2.5]{vershynin2018high}, which can be large for some $\ell$ when $\mbE(\|X_t-x_t\|)$ is large. 
\subsection{Averaged Moment Generating Function}
Motivated by the advantages and limitations of MGF, we resort to a modified version of MGF termed \textit{Averaged Moment Generating Function} (AMGF) in our method. 
\begin{definition}[AMGF]\label{def: salf}
    Given $\lambda\in\R$, the Averaged Moment Generating Function (AMGF) 
    is defined as
    \begin{align}\label{eq: AMGF}
        &\mbE_X(\Phi_{n,\lambda}(X))=\mbE_X\expectw{\ell\sim\mS^{n-1}}{e^{\lambda\innerp{\ell,X}}}.
    \end{align}
\end{definition}

The AMGF was recently proposed in \cite{altschuler2022concentration} in the field of Sampling Theory. It has a natural interpretation as the average of MGF over the unit sphere $\ell\in\mS^{n-1}$. AMGF can also be viewed as a special MGF while replacing the exponential energy function $e^{\lambda\innerp{\ell,x}}$ by its sphere-averaged version $\Phi_{n,\lambda}(x)=\expectw{\ell\sim\mS^{n-1}}{e^{\lambda\innerp{\ell,x}}}$. This energy function $\Phi_{n,\lambda}(x)$ exhibits several intriguing properties. 
\begin{lemma}[Properties of AMGF]\label{lemma: AMGF_properties}
    For $\salf{x}$ defined in \eqref{eq: AMGF}, we have 
    \begin{enumerate}
        \item {\it Rotation Invariance:} For any $x\in\R^n$ and any unitary matrix $Q\in\R^{n\times n}$, $$\salf{x}=\salf{Qx}.$$
        \item {\it Monotonicity:} For any $x,y\in\R^n$ such that $\|x\|\leq\|y\|$, 
        $$1\le\salf{x}\leq \salf{y}.$$
        \item {\it Decoupling:} Given $x\in\R^n$ and $w\sim subG(\sigma^2)$,
        \begin{equation}\label{eq: AMGF prop2}
            \mathbb{E}_w(\salf{x+w})\leq e^{\frac{\lambda^2\sigma^2}{2}}\salf{x}.
        \end{equation}
    \end{enumerate}
\end{lemma}
Moreover, AMGF induces the same norm concentration property as MGF.
\begin{lemma}\label{lemma: salf to sg}
    For a random variable $X\in\R^n$, if there exists $\sigma>0$ such that
    \begin{equation}\label{eq: salf nc}
    \mbE_x\left(\salf{x}\right)\leq e^{\frac{\lambda^2\sigma^2}{2}},~\forall \lambda\in\R,
\end{equation}
    then for any $\delta\in(0,1)$, \eqref{eq: concentration norm} holds with probability at least $1-\delta$.
\end{lemma}
We emphasize that even though \eqref{eq: salf nc} is a weaker condition than sub-Gaussian, it still guarantees norm concentration property as MGF. To see why Lemma \ref{lemma: salf to sg} holds, define an intermediate random variable $\tX=QX$ where $Q\sim\mathbb{U}^n$ is a random unitary matrix with $\mathbb{U}^n$ denoting the set of all the unitary matrices in $\R^{n\times n}$. Then the AMGF over $X$ is equal to the MGF over $\tX$, which means $\tX$ is sub-Gaussian with variance proxy $\sigma^2$ if \eqref{eq: salf nc} holds. Norm concentration of $X$ then follows by noticing that the transformation $\tX=QX$ does not affect the norm. 


\subsection{Theoretical Analysis}
Equipped with the AMGF, we establish a probabilistic bound of order $\mO(\sqrt{\log (1/\delta)})$ for the stochastic deviation $\|X_t-x_t\|$ by bounding the evolution of AMGF $\mbE(\salf{X_t-x_t})$ over time.

\begin{thm}\label{thm: high-prob}
     Consider the stochastic system \eqref{sys: d-t ss} and its associated deterministic system \eqref{sys: d-t ds} under Assumptions \ref{ass: Lipschitz f} and \ref{ass: bounded sigma}. Let $X_\tau$ be the trajectory of \eqref{sys: d-t ss} and $x_\tau$ be the associated trajectory of \eqref{sys: d-t ds} with the same initial state $x_0$ and input $u_\tau:\mathbb{Z}_{\ge 0}\to\mU$. Then, for any $t\geq0$, $\delta\in(0,1)$ and $\varepsilon\in(0,1)$,
     \begin{equation}\label{eq: thm1}
         \|X_t-x_t\|\leq\sqrt{\Psi_t(\varepsilon_1n+\varepsilon_2\log(1/\delta))}
     \end{equation}
     holds with probability at least $1-\delta$, where $\Psi_t$ is as in \eqref{eq: Psi val} and $\varepsilon_1,\varepsilon_2$ are as in \eqref{eq: epsilon val}.
\end{thm}

\begin{pf}
We begin with the special case where the Lipschitz constant $L=1$ at any time $t\geq0$. 
Define $v_t=X_t-x_t$ and $\beta_t=f(X_t,u_t,t)-f(x_t,u_t,t)$. By \eqref{sys: d-t ss} and \eqref{sys: d-t ds} we have
\begin{equation}
     v_{t+1}=\beta_t + w_t.
\end{equation}
By Assumption \ref{ass: Lipschitz f}, 
\begin{equation}\label{eq: v_t+1 d}
    \|\beta_t\|\leq \|v_t\|.
\end{equation}
By Assumption \ref{ass: bounded sigma} and Lemma \ref{lemma: AMGF_properties}(3), the conditional expectation of $\salf{v_{t+1}}$ can be bounded as
\begin{equation}\label{eq: x+w_t}
    \begin{split}
    &\mbE\left(\salf{v_{t+1}}|v_t\right)=\mbE_{w_t}\left(\salf{\beta_t+w_t}\right) \\
    &\leq e^{\frac{\lambda^2\sigma_t^2}{2}}\salf{\beta_t}
    \end{split}
\end{equation}
Combining \eqref{eq: v_t+1 d} and \eqref{eq: x+w_t}, invoking Lemma \ref{lemma: AMGF_properties}(2), we obtain
\begin{equation}\label{eq: salf on vt}
    \begin{split}
        &\expect{\Phi_{n,\lambda}(v_{t+1})|v_t}\leq e^{\frac{\lambda^2\sigma_t^2}{2}}\Phi_{n,\lambda}(v_t).
    \end{split}
\end{equation} 

Taking the expectation over $v_t$ on both sides of \eqref{eq: salf on vt} yields
\begin{equation*}\label{eq: E(AMGF)}
    \expect{\Phi_{n,\lambda}(v_{t+1})}\leq e^{\frac{\lambda^2\sigma_t^2}{2}}\mbE\left(\Phi_{n,\lambda}(v_t)\right),~ \expect{\Phi_{n,\lambda}(v_0)}=0. 
\end{equation*}
which is a linear difference inequality that points to
    \begin{equation}\label{eq: E(Phi)<=,L=1}
        \expect{\Phi_{n,\lambda}(v_t)}\leq e^{\frac{\lambda^2\sum_{k=0}^{t-1}\sigma_k^2}{2}}.
    \end{equation}
By Lemma \ref{lemma: salf to sg}, it follows that for any $\delta\in(0,1)$
    \begin{equation}\label{eq: result L=1}
         \|X_t-x_t\|\leq \sqrt{\sum_{k=0}^{t-1}\sigma_k^2(\varepsilon_1n+\varepsilon_2\log(1/\delta))}
     \end{equation}
holds with probability at least $1-\delta$, where $\varepsilon_1,\varepsilon_2$ are as in \eqref{eq: epsilon val}. This completes the proof of the special case.

Next, we consider the general cases under Assumption \ref{ass: Lipschitz f}. 
Define $\tX_t=\prod_{k=0}^{t-1} L_k^{-1}X_t$ and $\tx_t=\prod_{k=0}^{t-1} L_k^{-1}x_t$. Then $\tX_t$ satisfies 
\begin{equation}\label{sys: tx ss}
    \begin{split}
         \tX_{t+1}&=\prod_{k=0}^{t} L_k^{-1}f(\prod_{k=0}^{t-1} L_k\tX_t)+\prod_{k=0}^{t} L_k^{-1}w_t \\
         &\defeq g(\tX_t)+\tw_t
    \end{split}
\end{equation}
and similarly, $\tx_t$ satisfies 
\begin{equation}\label{sys: tx ds}
    \tx_{t+1}=\prod_{k=0}^{t} L_k^{-1}f(\prod_{k=0}^{t-1} L_k\tx_t)= g(\tx_t),
\end{equation}
which has the same drift term as \eqref{sys: tx ds}. For any $\tx_t,\ty_t\in\R^n$, it holds that
\begin{equation*}
    \begin{split}
        &\|g(\tx_t)-g(\ty_t)\|=\prod_{k=0}^{t} L_k^{-1}\|f(\prod_{k=0}^{t-1} L_k\tx_t)-f(\prod_{k=0}^{t-1} L_k\ty_t)\| \\
        &\leq \prod_{k=0}^{t} L_k^{-1}\cdot L_t\cdot \prod_{k=0}^{t-1} L_k\|\tx_t-\ty_t\|=\|\tx_t-\ty_t\|,
    \end{split}
\end{equation*}
meaning the scaled systems \eqref{sys: tx ss} and \eqref{sys: tx ds} satisfy Assumption \ref{ass: Lipschitz f} with $\tilde{L}=1$. Also, $\tw_t$ satisfies Assumption \ref{ass: bounded sigma} with variance proxy $\tilde{\sigma}^2_t=\prod_{k=0}^{t}L_k^{-2}\sigma_t^2$ due to the sum-up property of independent sub-Gaussian noise \cite{rigollet2023high}.
Thus the result of the special case can be applied. 

By \eqref{eq: result L=1}, with probability at least $1-\delta$,
    \[
        \|\tX_t-\tx_t\| \le \sqrt{\sum_{k=0}^{t-1}\sigma^2_{k}\psi_k^{-1}(\varepsilon_1n+\varepsilon_2\log(1/\delta))}.
    \]
Because of the relation $\|X_t-x_t\|=\prod_{k=0}^{t-1} L_k\|\tX_t-\tx_t\|$, the probabilistic bound \eqref{eq: thm1} follows. This completes the proof.
\end{pf}

\begin{remark}
    When Assumptions \ref{ass: Lipschitz f} and \ref{ass: bounded sigma} hold with time-invariant $L_t\equiv L$ and $\sigma_t\equiv\sigma$, $\psi_t$ defined in \eqref{eq: Psi val} becomes $\psi_t=L^{2t}$, and the result of Theorem \ref{thm: high-prob} reduces to
    \begin{equation}\label{eq:timeinvariant}
    \|X_t-x_t\|\leq \sqrt{\frac{\sigma^2(L^{2t}-1)}{L^2-1}(\varepsilon_1n+\varepsilon_2\log(1/\delta))}
\end{equation}
hold with probability at least $1-\delta$.
\end{remark}
\begin{remark}
In this work, we use Euclidean norm to measure the stochastic deviation $\|X_t-x_t\|$ for simplicity, but
the results in Theorem \ref{thm: high-prob} can be easily extended to the case with weighted Euclidean norm through a coordinate transformation \cite[Section V-D]{szy2024TAC}.
\end{remark}

Theorem \ref{thm: high-prob} establishes $r_{\delta,t}=\sqrt{\Psi_t(\varepsilon_1n+\varepsilon_2\log(1/\delta))}$ as a probabilistic bound on stochastic deviation. The bound is comprised of a scaling term $\sqrt{\Psi_t}$, a dimensional term $\sqrt{n}$ and a probabilistic term $\sqrt{\log (1/\delta)}$. The scaling term is determined by the Lipschitz constants $L_t$ and the disturbance variance proxy $\sigma_t^2$. It reveals the fluctuation of the system. Especially, when the system is linear, $\Psi_t$ is exactly the \textit{variance proxy} of $X_t-x_t$. The term $\sqrt{n}$ reveals the dimensional dependence of $r_{\delta,t}$, which is the same as the dimensional dependence of sub-Gaussian vectors shown in Lemma \ref{lemma: concentration}. The term $\sqrt{\log (1/\delta)}$ essentially represents the fast-decay feature of the distribution of $X_t-x_t$. Notice that this term means the probability level $\delta$ has an $\mO(e^{-r_{\delta,t}^2})$ dependence on the probabilistic bound $r_{\delta,t}$, so $\proba{\|X_t-x_t\|>r_{\delta,t}}$ decreases with $r_{\delta,t}$ even faster than the exponential rate. This phenomenon leads to the following question: Is this the best rate one can obtain? In other words, is $r_{\delta,t}$ derived from Theorem \ref{thm: high-prob} tight? 

 \subsection{Tightness of The Bound}\label{subsec: tightness}
 We next show that the probabilistic bound in Theorem \ref{thm: high-prob} is tight under Assumptions \ref{ass: Lipschitz f} and~\ref{ass: bounded sigma} and it is impossible to achieve better probabilistic bounds than \eqref{eq: thm1} without additional assumptions. In particular, we show that the tightness is precisely captured by linear systems satisfying Assumptions \ref{ass: Lipschitz f} and~\ref{ass: bounded sigma}.

Consider the linear stochastic system
$$x_{t+1}=Ax_t+Bu_t+w_t,~~w_t\sim subG(\sigma^2),$$ 
that satisfies Assumptions \ref{ass: Lipschitz f} and~\ref{ass: bounded sigma} with $L\equiv\|A\|$ and $\sigma_t\equiv\sigma$. Applying the time-invariant $L$ and $\sigma$ into \eqref{eq: Psi val}, we get that $\Psi_t=\frac{\sigma^2(L^{2t}-1)}{L^2-1}$. Therefore, by Theorem \ref{thm: high-prob}, we get that with probability at least $1-\delta$,
    \begin{equation}\label{eq: linear}
       \|X_t-x_t\|\leq\sqrt{\frac{\sigma^2(L^{2t}-1)}{L^2-1}(\varepsilon_1n+\varepsilon_2\log(1/\delta))}.
    \end{equation}
This is exactly the same as the $r_{\delta,t}^{(lin)}$ in \eqref{eq: lin sv bound}. From \cite[Chapter 4.6]{vershynin2018high}, we know that at least for linear systems satisfying Assumptions \ref{ass: Lipschitz f} and~\ref{ass: bounded sigma}, the dependence on $\delta$ and $n$ can not be improved further without additional assumptions. Therefore, $r_{\delta,t}$ is tight in our problem setting. 

\subsection{Comparison with Worst-Case Analysis}\label{subsec: comparison}
The worst-case analysis is a popular method for understanding the reachability when the disturbance is bounded \cite{cosner2024bounding}. We point out that it can also be applied to stochastic systems to obtain a probabilistic bound on the stochastic deviation under unbounded stochastic disturbance. As explained below, this is achieved by viewing unbounded sub-Gaussian disturbance as a bounded disturbance with high probability, reminiscent of conformal prediction \cite{hashemi2023data,vlahakis2024probabilistic}. However, the result is still significantly more conservative than our bound in Theorem \ref{thm: high-prob}. 

By norm concentration (Lemma \ref{lemma: concentration}) of sub-Gaussian random variables, for any $\delta\in(0,1)$,
\begin{equation}\label{eq: bound wt}
    \proba{\|w_\tau\|\leq \sqrt{\sigma_\tau^2(\varepsilon_1n+\varepsilon_2\log\frac{t}{\delta})}}\geq 1-\frac{\delta}{t}.
\end{equation}
Applying union bound to \eqref{eq: bound wt} over $\tau=0,\dots,t-1$, we obtain
\begin{equation}\label{eq: wt union}
    \proba{\|w_\tau\|\leq \sqrt{\sigma_\tau^2(\varepsilon_1n+\varepsilon_2\log\frac{t}{\delta})},~\forall \tau<t}\geq 1-\delta.
\end{equation}
This means that the disturbance is bounded with
    \begin{equation}\label{eq: b val}
    \|w_\tau\|\le b_\tau:= \sqrt{\sigma_\tau^2(\varepsilon_1n+\varepsilon_2\log\frac{t}{\delta})}
\end{equation}
for all $\tau< t$ with probability at least $1-\delta$.
A worst-case probabilistic bound on $\|X_t-x_t\|$ can be established by assuming this bound. More specifically, by the system dynamics \eqref{sys: d-t ss}-\eqref{sys: d-t ds} and the triangular inequality,
\begin{equation}
\begin{split}
    \|X_{\tau+1}-x_{\tau+1}\|\leq&\|f(X_\tau,u_\tau,\tau)-f(x_\tau,u_\tau,\tau)\|+\|w_\tau\| \\
    \leq& L_\tau\|X_\tau-x_\tau\|+b_\tau 
\end{split}
\end{equation}
for all $\tau< t$ with probability at least $1-\delta$.
It follows that
\begin{equation}\label{eq: xt-xt worst}
    \|X_t-x_t\|\leq \sqrt{\psi_{t-1}}\sum_{k=0}^{t-1}b_{k}\sqrt{\psi_k^{-1}},
\end{equation}
where $\psi_t$ is as in \eqref{eq: Psi val}. Plugging \eqref{eq: b val} into \eqref{eq: xt-xt worst}, we conclude that, for any $\delta\in(0,1)$,
\begin{equation}\label{eq: sd by worst}
    \|X_t-x_t\|\leq \sqrt{\psi_{t-1}}\sum_{k=0}^{t-1}\sigma_{k}\sqrt{\psi_k^{-1}(\varepsilon_1n+\varepsilon_2\log\frac{t}{\delta})}.
\end{equation}
holds with probability at least $1-\delta$. 


This bound \eqref{eq: sd by worst} derived using worst-case analysis is substantially more conservative than that in Theorem \ref{thm: high-prob}. First, since $\sqrt{\Psi_t}\leq \sqrt{\psi_{t-1}}\sum_{k=0}^{t-1}\sigma_{k}\sqrt{\psi_k^{-1}}$, \eqref{eq: sd by worst} is always worse than \eqref{eq: thm1}.
To see more clearly the gap, consider the case when $L_t\equiv L$ and $\sigma_t\equiv\sigma$. In this case, \eqref{eq: sd by worst} reduces to $\|X_t-x_t\|\leq \frac{L^t-1}{L-1}\sqrt{\sigma^2(\varepsilon_1n+\varepsilon_2\log\frac{t}{\delta})}$, which is much worse than \eqref{eq:timeinvariant} when $L$ is close to 1. In particular,
when $L=1$, $\sqrt{\psi_t}\sum_{k=1}^t\sigma_{k-1}\sqrt{\psi_t^{-1}}=\sigma t$, significantly larger than $\sqrt{\Psi_t}=\sigma\sqrt{t}$ as $t$ increases. 

\section{Probabilistic Reachable Set}\label{sec: PRS}
With the tight probabilistic bound \eqref{eq: thm1} on stochastic deviation, we can integrate it with any existing methods for approximating the DRS for \eqref{sys: d-t ds}, by leveraging the separation strategy outlined in Proposition \ref{prop: separation}, to obtain a practical $\delta$-PRS for \eqref{sys: d-t ss}.


\begin{thm}\label{thm: reachable set}
    Consider the stochastic system \eqref{sys: d-t ss} with the initial set  $\mathcal{X}_0\subseteq \R^n$ and the input set $\mathcal{U}\subseteq \R^p$. Suppose that Assumptions \ref{ass: Lipschitz f} and \ref{ass: bounded sigma} hold. Then given any over-approximation $\overline{\mR}_t$ of the DRS of the deterministic system \eqref{sys: d-t ds}, $\mR_{\delta,t}=\overline{\mR}_t\oplus\mathcal{B}^n\left(r_{\delta,t},0\right)$
   is a $\delta$-PRS of the system \eqref{sys: d-t ss}, where
$r_{\delta,t}=\sqrt{\Psi_t(\varepsilon_1n+\varepsilon_2\log(1/\delta))}$ with $\Psi_t$ as \eqref{eq: Psi val} and $\varepsilon_1,\varepsilon_2$ as \eqref{eq: epsilon val}. 
\end{thm}

\begin{pf}
    Plug the bound stated in Theorem \ref{thm: high-prob} into \eqref{eq: lemma traj2set} in Proposition \ref{prop: separation}, then Theorem \ref{thm: reachable set} follows. 
\end{pf}

Since $r_{\delta,t}=\sqrt{\Psi_t(\varepsilon_1n+\varepsilon_2\log(1/\delta))}$ is proved to have tight dependence on the probability level $1-\delta$, $\mR_{\delta,t}$ in Theorem \ref{thm: reachable set} is also tight with respect to $\delta$. In scenarios like safety-critical control, $\mR_{\delta,t}$ only scales at the rate of $\mO(\sqrt{\log (1/\delta)})$, and is thus suitable for harsh probability-level requirements that need an extremely small $\delta$. As the dependence of $r_{\delta,t}$ on $\delta$ and $n$ cannot be further improved, the tightness of $\mR_{\delta,t}$ merely depends on the tightness of the over-approximation $\overline{\mR}_t$ on the DRS. It becomes loose when $\overline{\mR}_t$ is conservative. 

Besides tightness, computational efficiency is also an important consideration. The computational cost of $\mR_{\delta,t}$ in Theorem \ref{thm: reachable set} arises from two main sources: computing $\overline{\mR}_t$ and realizing the Minkowski sum $\oplus$ with a ball. Although computing the Minkowski sum in a parameterized form is challenging and efficient algorithms are only available for ellipsoids and polyhedra \cite{varadhan2004accurate,weibel2007minkowski}, in practice we only need an efficient membership oracle to determine whether $X_t\in \overline{\mR}_t\oplus\mathcal{B}^n\left(r_{\delta,t},0\right)$. This can be achieved by checking $\min_{y\in\overline{\mR}_t} \|y-X_t\|\leq r_{\delta,t}$, which is simpler and is a convex optimization when $\overline{\mR}_t$ is convex. Therefore, the computational efficiency primarily depends on how $\overline{\mathcal{R}}_t$ is generated.
 
Following the above discussion on the tightness and computational efficiency, we conclude that if the reachability of the associated deterministic system \eqref{sys: d-t ds} is well studied, then $\delta$-PRS can be easily obtained by Theorem \ref{thm: reachable set}.  In the following subsections, we exemplify this conclusion with two popular types of reachability study for deterministic systems. These methods are scalable and result in convex $\overline{\mR}_t$, rendering efficient algorithms for probabilistic reachability analysis.  

\subsection{Lipschitz Bound Reachability}
In this section, we review a computationally efficient method for reachability using Lipschitz bound of the dynamics~\cite{ZH-SM:12}. We start with an assumption. 
\begin{assumption}\label{assum:1}
    For system~\eqref{sys: d-t ds}, there exist constants $L_d,\rho\ge 0$ such that, for every $x,u,t \in \R^n\times \mathcal{U}\times \mathbb{Z}_{\ge 0}$,
    \begin{enumerate}
        \item $\|D_x f(x,u,t)\|_{\mathbb{X}}\le L_d$, and 
        \item $\|D_u f(x,u,t)\|_{\mathbb{X},\mathbb{U}}\le \rho$,
    \end{enumerate}
    where $\|\cdot\|_{\mathbb{X}}$ is a norm on $\R^n$, $\|\cdot\|_{\mathbb{U}}$ is a norm on $\R^p$, and $\|\cdot\|_{\mathbb{X},\mathbb{U}}$ is the induced norm on $\R^{n\times p}$. 
\end{assumption}
The norms $\|\cdot\|_{\mathbb{X}}$ and$\|\cdot\|_{\mathbb{U}}$ can be chosen different from Euclidean norm to ensure the tightest over-approximation of the reachable sets of the deterministic system~\eqref{sys: d-t ds}. 
%
Suppose that Assumption~\ref{assum:1} holds for the discrete-time system~\eqref{sys: d-t ds} and $t\mapsto x^*_t$ is a trajectory of~\eqref{sys: d-t ds} with an input $t\mapsto u^*_t$. Define $\mathcal{B}_{\mathbb{X}}(r,y)=\{x\in\R^n: \|x-y\|_{\mathbb{X}}\leq r\}$. Then, for every $x_0\in \mathcal{B}_{\mathbb{X}}(r_1,x^*_0)$ and every input $t\mapsto u_t$ with $u_t\in \mathcal{B}_{\mathbb{U}}(r_2,u^*_t)$, we have 
\begin{align}\label{eq:over-c}
    \overline{R}_t = \mathcal{B}_{\mathbb{X}}\left(L_d^t r_1 + \rho (\tfrac{L_d^{t-1}-1}{L_d-1}) r_2,x^*_t\right). 
\end{align}
We can use the Lipschitz bound over-approximation~\eqref{eq:over-c} to construct $\delta$-PRS for the stochastic system~\eqref{sys: d-t ss}. 

\begin{proposition}[Lipschitz bound Reachability] \label{prop: case 1}
Consider the stochastic system~\eqref{sys: d-t ss} with the associated deterministic system~\eqref{sys: d-t ds} satisfying Assumptions~\ref{ass: Lipschitz f}, \ref{ass: bounded sigma}, and~\ref{assum:1}. Let $t\mapsto x^*_t$ be a trajectory of~\eqref{sys: d-t ds} with an input $t\mapsto u^*_t$ and $t\mapsto X_t$ be a trajectory of~\eqref{sys: d-t ds} starting from $x_0\in \mathcal{B}_{\mathbb{X}}(r_1,x^*_0)$ with an input $t\mapsto u_t\in \mathcal{B}_{\mathbb{U}}(r_2,u^*_t)$. Then, with probability at least $1-\delta$, 
\begin{align*}
    X_t \in \mathcal{B}_{\mathbb{X}}\left(L_d^{t}r_1 + \rho (\tfrac{L_d^{t-1}-1}{L_d-1})r_2, x^*_t\right) \oplus \mathcal{B}^n\left(r_{\delta,t},0\right) 
\end{align*}
 where $r_{\delta,t}=\sqrt{\Psi_t(\varepsilon_1n+\varepsilon_2\log(1/\delta))}$ with $\Psi_t$ as \eqref{eq: Psi val} and $\varepsilon_1,\varepsilon_2$ as \eqref{eq: epsilon val}. 
\end{proposition}

\begin{pf}
    The proof is a straightforward combination of Lipschitz bound over-approximation~\eqref{eq:over-c} and Theorem~\ref{thm: reachable set}.  
\end{pf}

\subsection{Interval-Based Reachability}
Interval analysis provides a framework for estimating the propagation of uncertainties in discrete-time systems~\cite{LJ-MK-OD-EW:01}. 
The main idea of interval-based reachability is to embed the dynamical system into a higher dimensional space using a suitable inclusion function.
 The map $\left[\begin{smallmatrix}\underline{\OF}\\ \overline{\OF}\end{smallmatrix}\right]: \R^{2n}\times \R^{2p}\times \mathbb{Z}_{\ge 0}\to \R^{2n}$ is an inclusion function for $f$, if, for every $z,w\in [\underline{x},\overline{x}]\times [\underline{u},\overline{u}]$ and every $t\in \mathbb{Z}_{\ge 0}$, 
 \begin{align*}
     \underline{\OF}(\underline{x},\overline{x},\underline{u},\overline{u},t) \le f(z,w,t) \le \overline{\OF}(\underline{x},\overline{x},\underline{u},\overline{u},t). 
 \end{align*}
 Many automated approaches exist for finding an inclusion function for $f$. We refer to~\cite[Section IV.B]{SJ-AH-SC:23} for a detailed discussion on these approaches and to~\cite{harapanahalli2023toolbox} for a toolbox for computing inclusion functions.  
 Given an interval initial configuration $\mathcal{X}_0=[\underline{x}_0,\overline{x}_0]$ and an interval input set $\mathcal{U}=[\underline{u},\overline{u}]$, the embedding system of~\eqref{sys: d-t ds} associated with the inclusion function $\OF$ is given by
\begin{align}\label{eq:embedding}
  \begin{bmatrix}
      \underline{x}_{t+1}\\
      \overline{x}_{t+1}
  \end{bmatrix} = \begin{bmatrix}
      \underline{\OF}(\underline{x}_t,\overline{x}_t,\underline{u},\overline{u},t)\\
      \overline{\OF}(\underline{x}_t,\overline{x}_t,\underline{u},\overline{u},t)
  \end{bmatrix}.  
\end{align}
Let $\left[\begin{smallmatrix}\underline{x}_t\\ \overline{x}_t\end{smallmatrix}\right]$ be the trajectory of the embedding system~\eqref{eq:embedding} starting from  $\left[\begin{smallmatrix}
    \underline{x}_0\\ \overline{x}_0
\end{smallmatrix}\right]$. Then, an over-approximation of the deterministic reachable set of~\eqref{sys: d-t ds} is
\begin{align}\label{eq:overapprox-interval}
    \overline{\mathcal{R}}_t= [\underline{x}_t,\overline{x}_t].
\end{align}
This over-approximation of reachable sets can be used in Theorem~\ref{thm: reachable set} to construct $\delta$-PRS of the system~\eqref{sys: d-t ss}.

\begin{proposition}[Interval-based Reachability]\label{prop: case 2}
    Consider the stochastic system~\eqref{sys: d-t ss} with the associated deterministic system~\eqref{sys: d-t ds} satisfying Assumptions~\ref{ass: Lipschitz f} and \ref{ass: bounded sigma}. Let $t\mapsto X_t$ be a trajectory of the stochastic system~\eqref{sys: d-t ss} starting from $x_0\in [\underline{x}_0,\overline{x}_0]$ with an input $t\mapsto u_t \in [\underline{u},\overline{u}]$ and $\left[\begin{smallmatrix}\underline{\OF}\\ \overline{\OF}\end{smallmatrix}\right]$ be an inclusion function for $f$, Then with probability at least $1-\delta$, 
    \begin{align*}
    X_t \in [\underline{x}_t,\overline{x}_t] \oplus \mathcal{B}^n\left(r_{\delta,t},0\right) 
\end{align*}
 where $\left[\begin{smallmatrix}\underline{x}_t\\ \overline{x}_t\end{smallmatrix}\right]$ is the trajectory of the embedding system~\eqref{eq:embedding} starting from  $\left[\begin{smallmatrix}
    \underline{x}_0\\ \overline{x}_0
\end{smallmatrix}\right]$ and $r_{\delta,t}=\sqrt{\Psi_t(\varepsilon_1n+\varepsilon_2\log(1/\delta))}$ with $\Psi_t$ as \eqref{eq: Psi val} and $\varepsilon_1,\varepsilon_2$ as \eqref{eq: epsilon val}. 
\end{proposition}

\begin{pf}
    The proof is a straightforward combination of interval-case over-approximation~\eqref{eq:overapprox-interval} and Theorem~\ref{thm: reachable set}.  
\end{pf}

\section{Numerical experiments}\label{sec: simulations}
\subsection{Linear System}
We first consider a linear example to validate the tightness of our bound \eqref{eq: thm1} on the stochastic deviation. Consider a simple linear dynamics
\begin{equation}\label{sys: linear}
    \begin{split}
        X_{t+1}&=-0.93I_nX_tdt+ w_t\\
        &\defeq AX_t+w_t,
    \end{split}
\end{equation}
with $w_t\sim N(0,0.2I_n)$ and initialized at $X_0=0$. The system \eqref{sys: linear} satisfies Assumption \ref{ass: Lipschitz f} with $L_t\equiv L = \|A\|=0.93$ and Assumption \ref{ass: bounded sigma} with $\sigma_t\equiv\sigma=\sqrt{0.2}$. By linearity, $X_t$ follows a zero-mean Gaussian distribution whose covariance $\text{cov}(X_t)$ can be computed using the sum-up property of zero-mean Gaussian variables. The trajectory of the deterministic dynamics associated with \eqref{sys: linear} starting from $x_0=0$ is $x_t\equiv 0$.

\begin{figure}[t]
	\centering
  \includegraphics[width =0.48\linewidth]{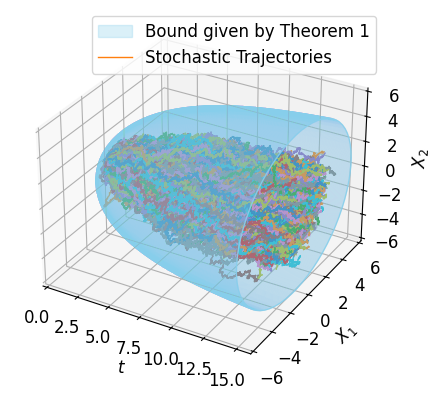}
    \includegraphics[width =0.48\linewidth]{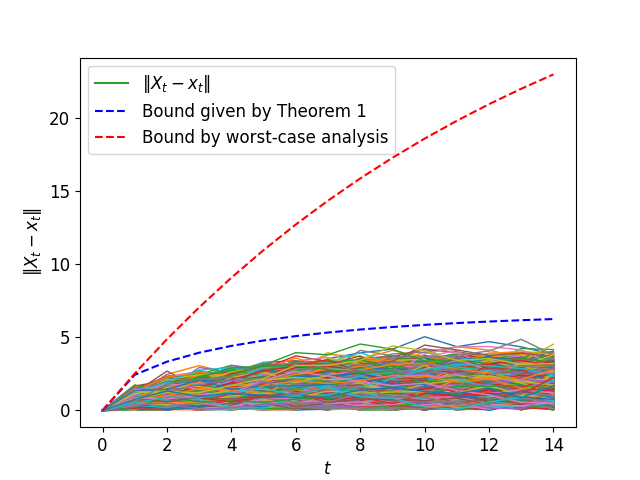}
	\caption{Probabilistic bound of stochastic deviation for a linear system. \textbf{Left:} each curve represents an independent trajectory of $X_t$ The radius of the blue envelope at time $t$ is the bound \eqref{eq: thm1}. \textbf{Right:} each solid curve is an independent trajectory of $\|X_t-x_t\|$. The blue dashed curve is the bound \eqref{eq: thm1}. The red dashed line is the bound \eqref{eq: sd by worst}.
 }
	\label{fig: Lin}
\end{figure}

To illustrate the bound \eqref{eq: thm1}, we simulate 5000 independent trajectories of \eqref{sys: linear} with $n=2$ over the time horizon $t\leq 15$ and compute the deviation associated with each trajectory, as depicted in Figure \ref{fig: Lin}. These trajectories are compared with our probabilistic bound \eqref{eq: thm1} with $\delta=10^{-3}$ and $\varepsilon=1/16$. Figure \ref{fig: Lin} shows that all the trajectories satisfy the bound $r_{\delta,t}$ as expected. Moreover, the bound \eqref{eq: sd by worst} derived by the worst-case analysis is put in the right sub-figure of Figure \ref{fig: Lin} to compare with our probabilistic bound \ref{eq: thm1}. It is clear that worst-case analysis produces a much more conservative probabilistic bound on stochastic deviation.  

\begin{figure}[t]
	\centering
         \includegraphics[width =0.48\linewidth]{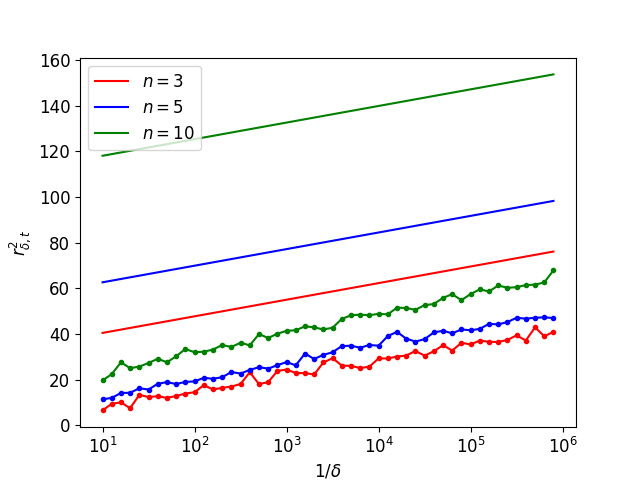}
  \includegraphics[width =0.48\linewidth]{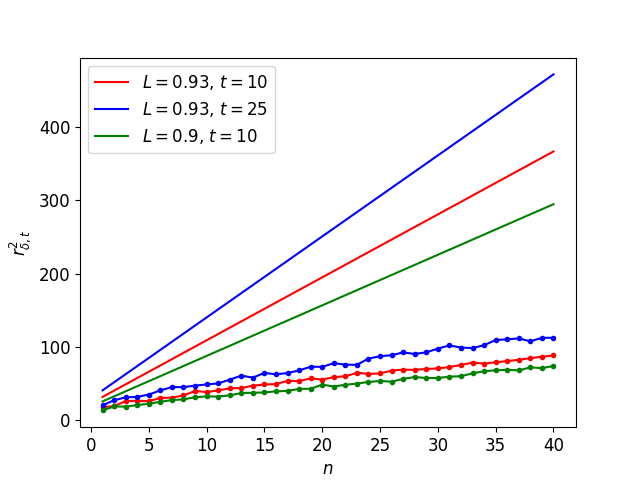}
	\caption{Illustration of the tightness of $r_{\delta,t}$ w.r.t. $\delta,n$. \textbf{Left:} the solid line shows the dependence of $r_{\delta,t}^2$ over $1/\delta$ and the dotted line in the same color is the corresponding simulated bound $\hat{r}_{\delta,t}^2$. The time is fixed as $t=25$. \textbf{Right:} the solid line shows the dependence of $r_{\delta,t}^2$ over $n$, and the dotted line in the same color is the corresponding simulated $\hat{r}_{\delta,t}^2$. $\delta=10^{-4}$ is fixed.
 }
	\label{fig: r-delta r-n}
\end{figure}

By Theorem \ref{thm: high-prob}, the square of our bound \eqref{eq: thm1}, denoted as $r_{\delta,t}^2$, grows linearly with $\log(1/\delta)$ and $n$, as illustrated in Figure \ref{fig: r-delta r-n}. To verify the tightness of these dependencies, we compare them with those obtained through Monte-Carlo simulation. In particular, for each choice of $\delta$ and $n$, we simulate $10^7$ independent trajectories of \eqref{sys: linear} and compute the associated value of $\|X_t-x_t\|$ for each trajectory. We follow a standard approach \cite{shapiro2003monte} and estimate the high probability bound $\hat{r}_{\delta,t}$ of the stochastic deviation as the $\delta$-th largest $\|X_t-x_t\|$ (e.g., top 1\% if $\delta=10^{-2}$). The results, shown in Figure \ref{fig: r-delta r-n}, imply that $\hat{r}_{\delta,t}^2$ also grows linearly with $\log(1/\delta)$ and $n$, consistent with our theoretical bound \eqref{eq: thm1}. 


\subsection{Cobweb Supply-Demand Model}
We next consider a Cobweb Supply-Demand model \cite{pindyck2012microeconomics} with nonlinear dynamics. This model describes the price-quantity relationship of an item in a nearly-saturated single market such as electronics \cite{hommes1994dynamics}. For a certain item, the model is described as 
\begin{equation}\label{eq: eco}
\begin{split}
    p_{t+1} &= a - b\log(1+q_t) +w_{t,1} \\
    q_{t+1} &= cp_{t+1}-d + w_{t,2}
\end{split}
\end{equation}
where $p_t$ is the price, $q_t$ is the quantity of the item, $a,~b,~c,~d$ are parameters determined by the market, and $w_{t,1},~w_{t,2}$ are disturbance brought by the market volatility. The model in the form of \eqref{sys: d-t ss} is
\begin{equation}\label{eq: eco sys}
    \begin{split}
        X_{t+1} = &\begin{bmatrix}
            a - b\log(1+q_t) \\
            (ca-d) - cb\cdot\log(1+q_t)
        \end{bmatrix} + \begin{bmatrix}
            w_{t,1} \\
            cw_{t,1}+w_{t,2}
        \end{bmatrix} \\
        \defeq &f(X_t)+w_t
    \end{split}
\end{equation}
where $X_t=[p_t ~q_t]^{\top}$. We set the parameters $a=10$, $b=1.5$, $c=0.5$, $d=1$ and $w_{t,i}$ as truncated Gaussian noise:
$w_{t,1} = \min\nolimits_{|\cdot|}\{N(0,10^{-5}), 0.05p_t\}$ and $w_{t,2} = \min\nolimits_{|\cdot|}\{N(0,10^{-5}), 0.05q_t\}$
where $\min_{|\cdot|}\{x, y\}$ outputs the one with smaller absolute value. The Lipschitz constant can be calculated as $L_t=\frac{b\sqrt{1+c^2}}{1+q_t}\in(0,1)$, and $\sigma_t\approx0.0032$ is approximated 
through Monte-Carlo simulation \cite[Chapter 2.5]{vershynin2018high}.

\begin{figure}[t]
 \centering
  \includegraphics[width =0.49\linewidth]{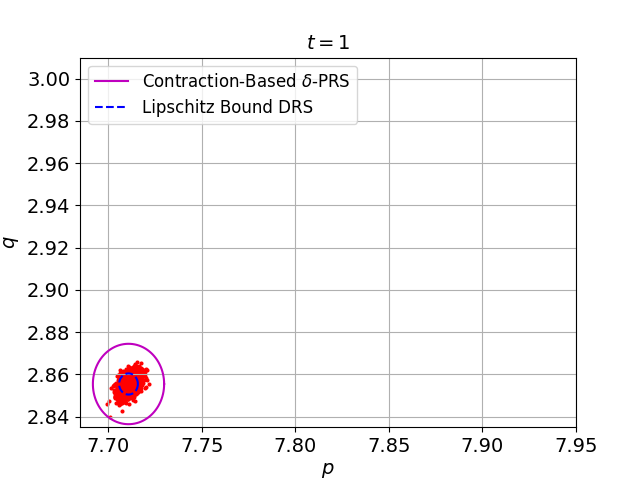}
  \includegraphics[width =0.49\linewidth]{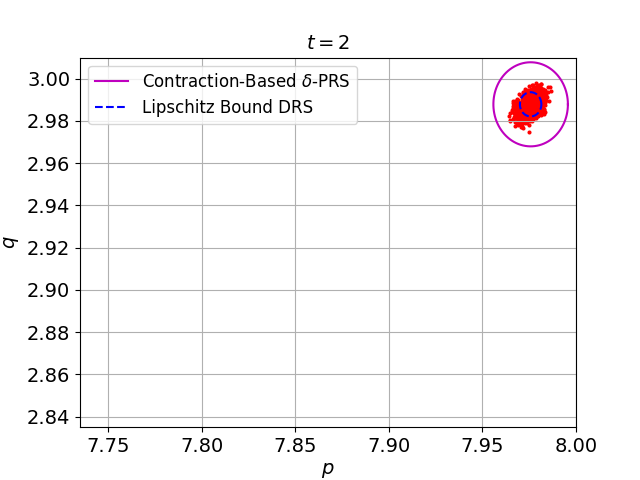}
    \includegraphics[width =0.49\linewidth]{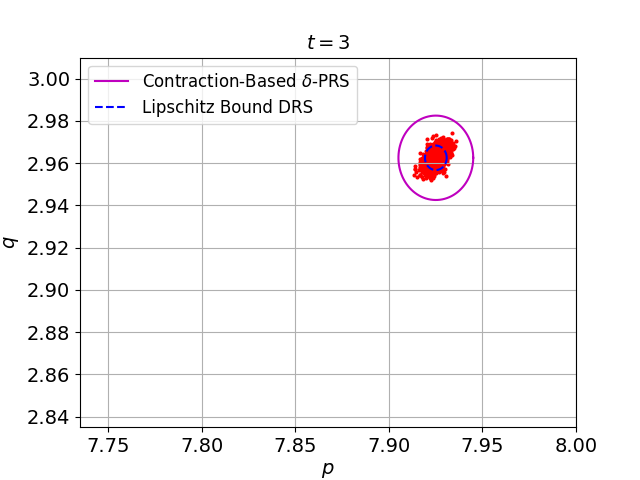}
  \includegraphics[width =0.49\linewidth]{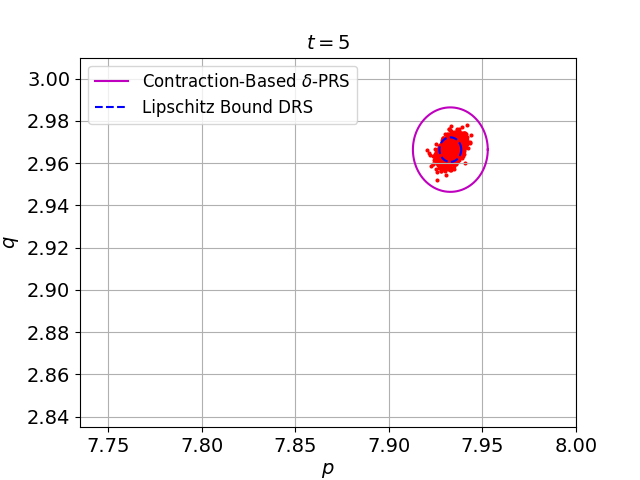}
  \caption{Evolution of the 2000 independent trajectories of the system \eqref{eq: eco sys} and the $\delta$-PRS at $t=1$ \textbf{(upper-left)}, $t=2$ \textbf{(upper-right)}, $t=3$ \textbf{(lower-left)} and $t=5$ \textbf{(lower-right)}. The purple circles are $\delta$-PRS calculated by Proposition \ref{prop: case 1}. The blue dashed circles are Lipschitz-bound DRS. The red points are the state of the trajectories at different times.}
	\label{fig:Cob}
 \end{figure}

We validate our $\delta$-PRS based on the Lipschitz bound reachability with the initial set $\mX_0=[9.195,9.205]\times[3.595,3.605]$. In this case, $x_t^{*}=(p_t^{*},q_t^{*})$ is the trajectory of the associated deterministic system of \eqref{eq: eco sys} starting from $x_0^{*}=[9.2~ 3.6]^{\top}$, and $r_1=5\sqrt{2}\times10^{-3}$. In each experiment, we sample 2000 independent trajectories starting from $X_0\in\mX_0$ during $t\leq5$, and the $\delta$-PRS based on Proposition \ref{prop: case 1}, with the norm $\|\cdot\|_{\mathbb{X}}=\|\cdot\|$, the probability level $\delta=10^{-3}$ and the parameter $\varepsilon=1/32$. The evolution of the stochastic trajectories and the $\delta$-PRS are visualized in Figure \ref{fig:Cob}. It is clear that all the sampled points are within the derived bound, thus validating the derived $\delta$-PRS.


\subsection{Fixed-Wing UAV Path Following}
In this section we consider the path following task of a discre-time fixed-wing  unmanned aerial vehicle (UAV), whose kinematic model is formulated as 
\begin{equation}\label{sys: UAV}
    \begin{split}
        X_{t+1}&=X_t+\eta\begin{bmatrix}
           v\cos\theta_t\cos\gamma_t + u_{t,x} \\
           v\sin\theta_t\cos\gamma_t + u_{t,y} \\
           v\sin\gamma_t + u_{t,z}\\
           \frac{g}{v}\tan\varphi_t
        \end{bmatrix}+w_t\\
        &\defeq f(X_t,u_t)+w_t,
    \end{split}
\end{equation}
where $X_t=[p_{t,x}~ p_{t,y}~ p_{t,z}~ \theta_t]^{\top}$ is the state variable at time $t$, $(p_{t,x},p_{t,y})$ is the inertial north-east position of the UAV, $p_{t,z}$ is the altitude, $\theta_t$ is the heading angle measured from north, $\eta$ is the discretization stepsize, $[\gamma_t~ \varphi_t]^{\top}$ is the controller input, $\gamma_t$ is the air mass referenced flight path angle, $\varphi_t$ is the roll angle, $v$ is the airspeed, $g$ is the gravity, $u_t=[u_{t,x}~ u_{t,y}~ u_{t,z}]^{\top}$ is the bounded noise brought by the wind, and $w_t$ is the stochastic disturbance in the environment. We set $v=13$, $g=9.8$, $\eta=0.1$, $w_t\sim \sqrt{\eta}\cdot 0.1N(0, diag([1,1,1,0.1]))$ and $|u_{t,i}|\leq 0.5$ for $i=x,y,z$. We use the feedback controller proposed in \cite[Section 3]{beard2014fixed} with the same feedback gains
to control the UAV \eqref{sys: UAV} starting from $X_0=[5~ 4.5~ 0~ \frac{5\pi}{18}]^{\top}$ to follow the line $l^*=[0~ 0~ 3]^{\top}+\alpha[1~ 1~ 0]^{\top}$, $\alpha\in[0,+\infty)$ in the $p_x-p_y-p_z$ space.

\begin{figure}[t]
 \centering
  \includegraphics[width =0.49\linewidth]{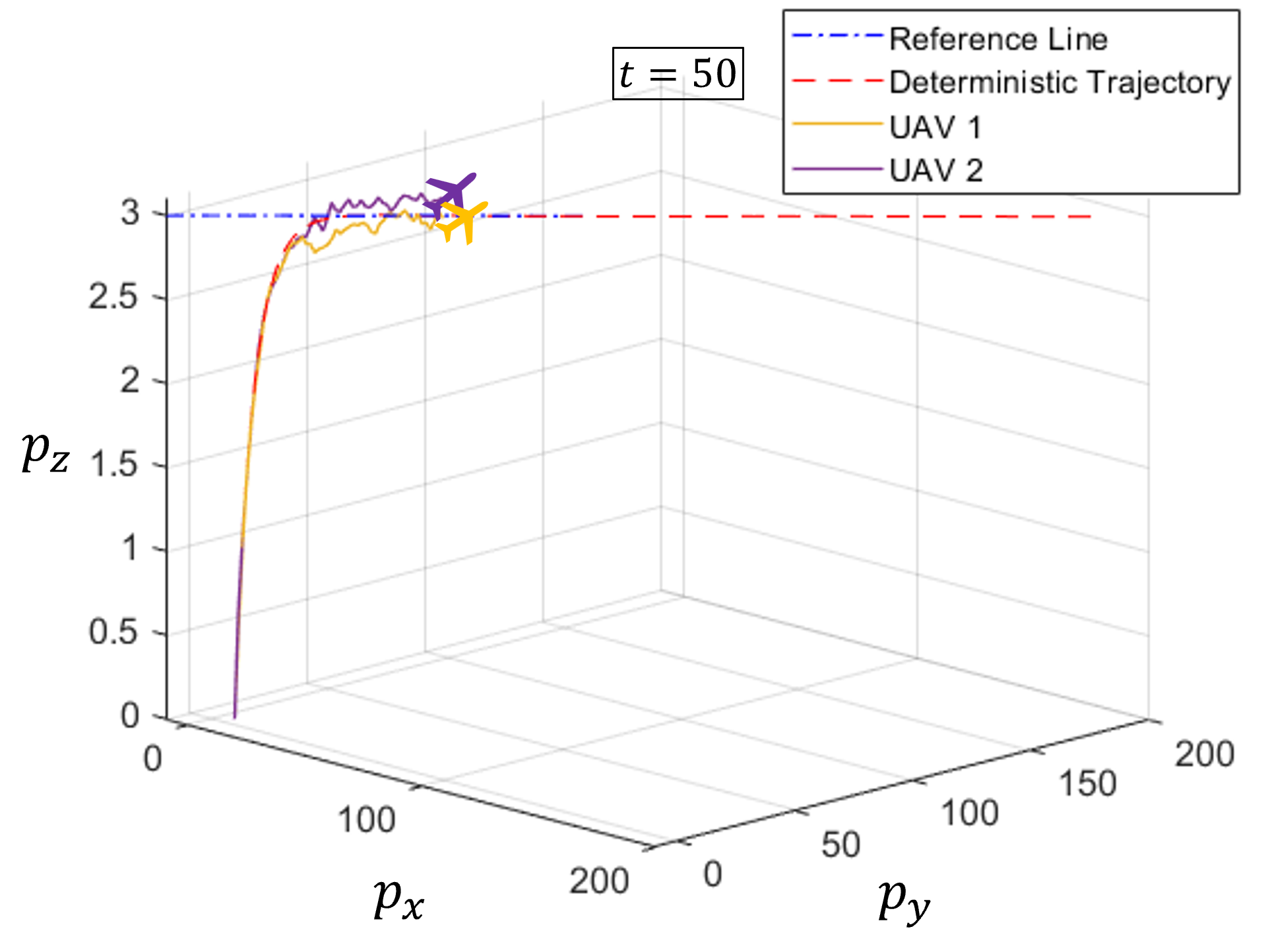}
  \includegraphics[width =0.49\linewidth]{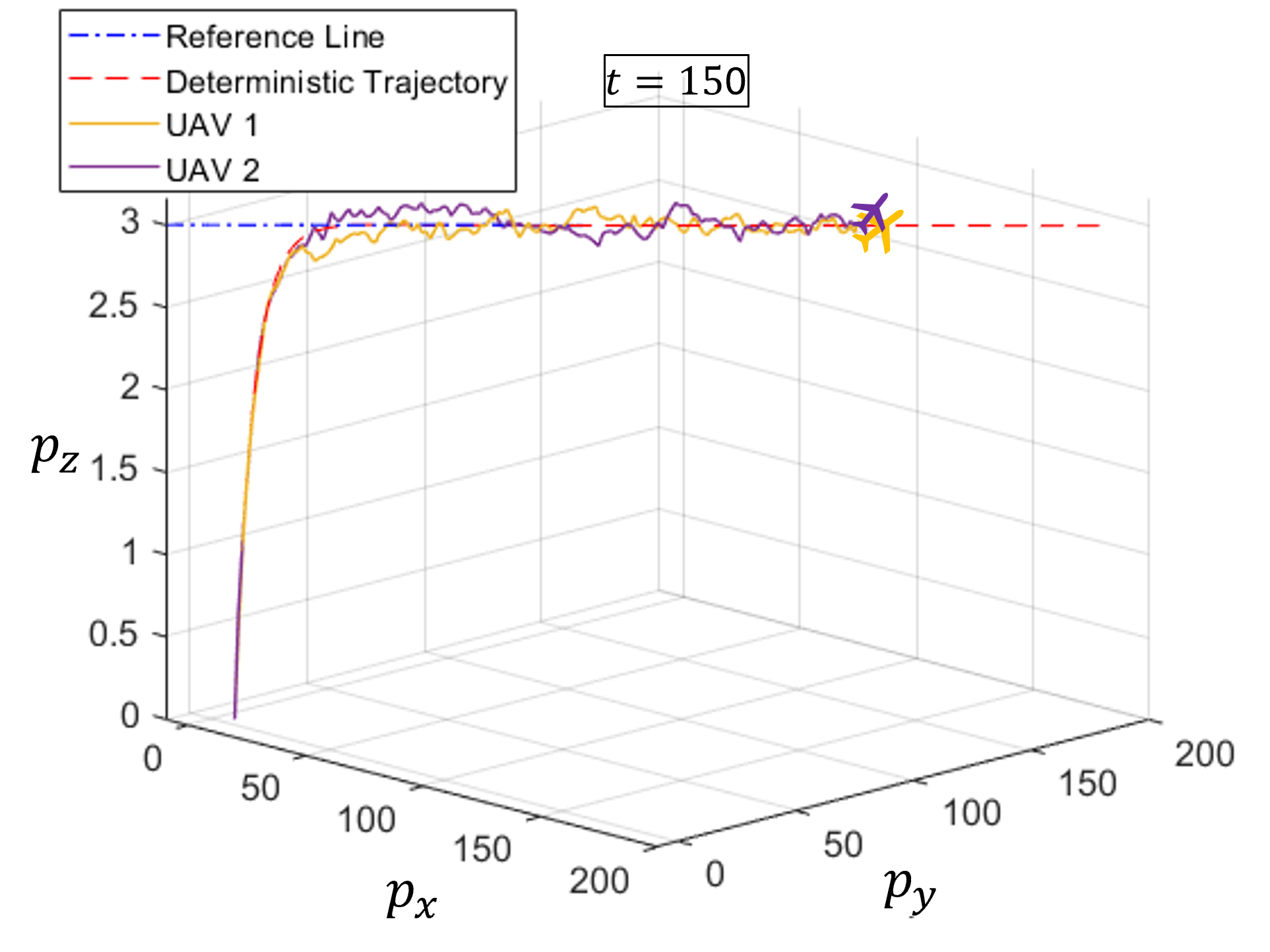}
  \includegraphics[width =0.49\linewidth]{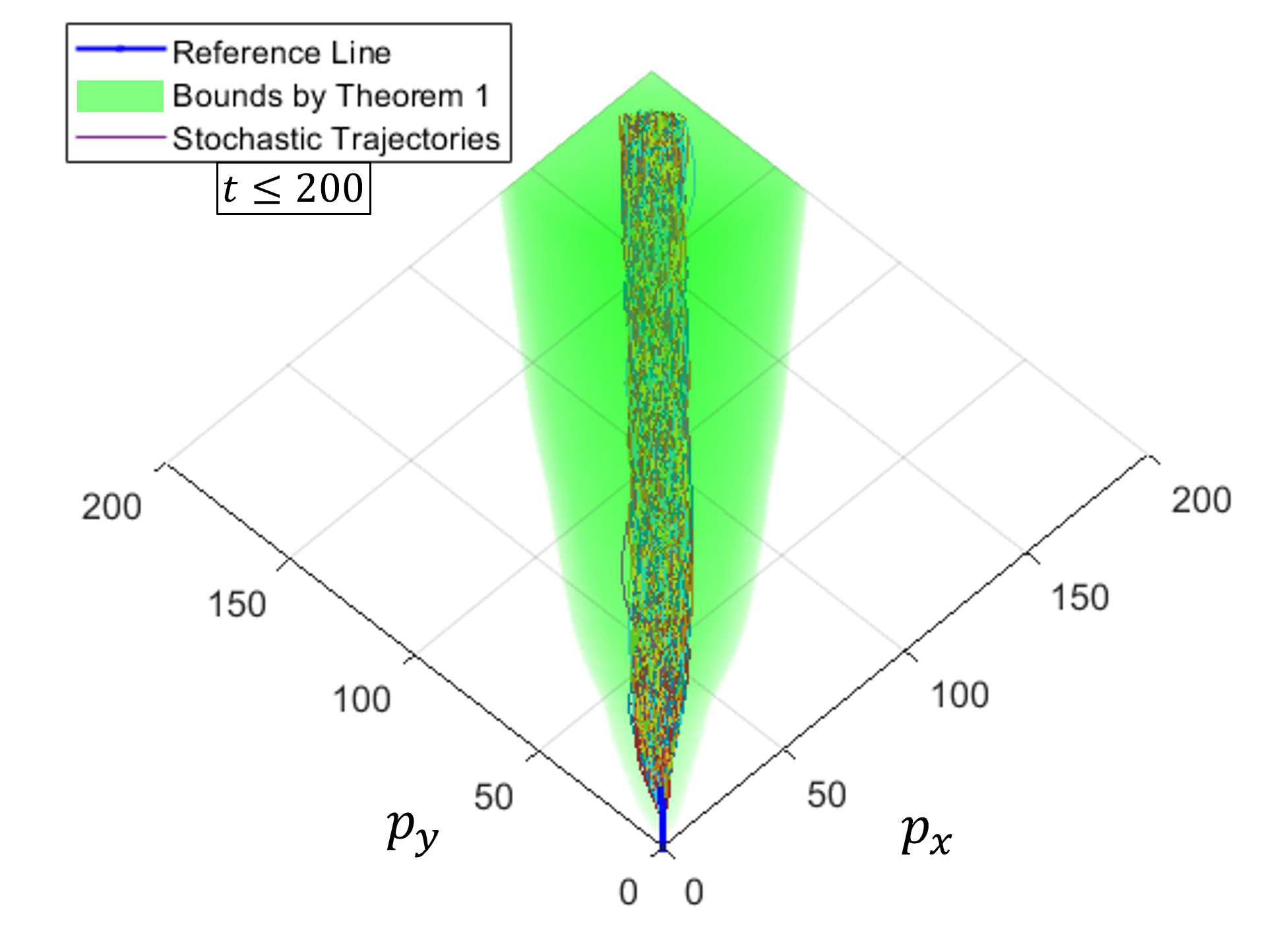}
  \includegraphics[width =0.49\linewidth]{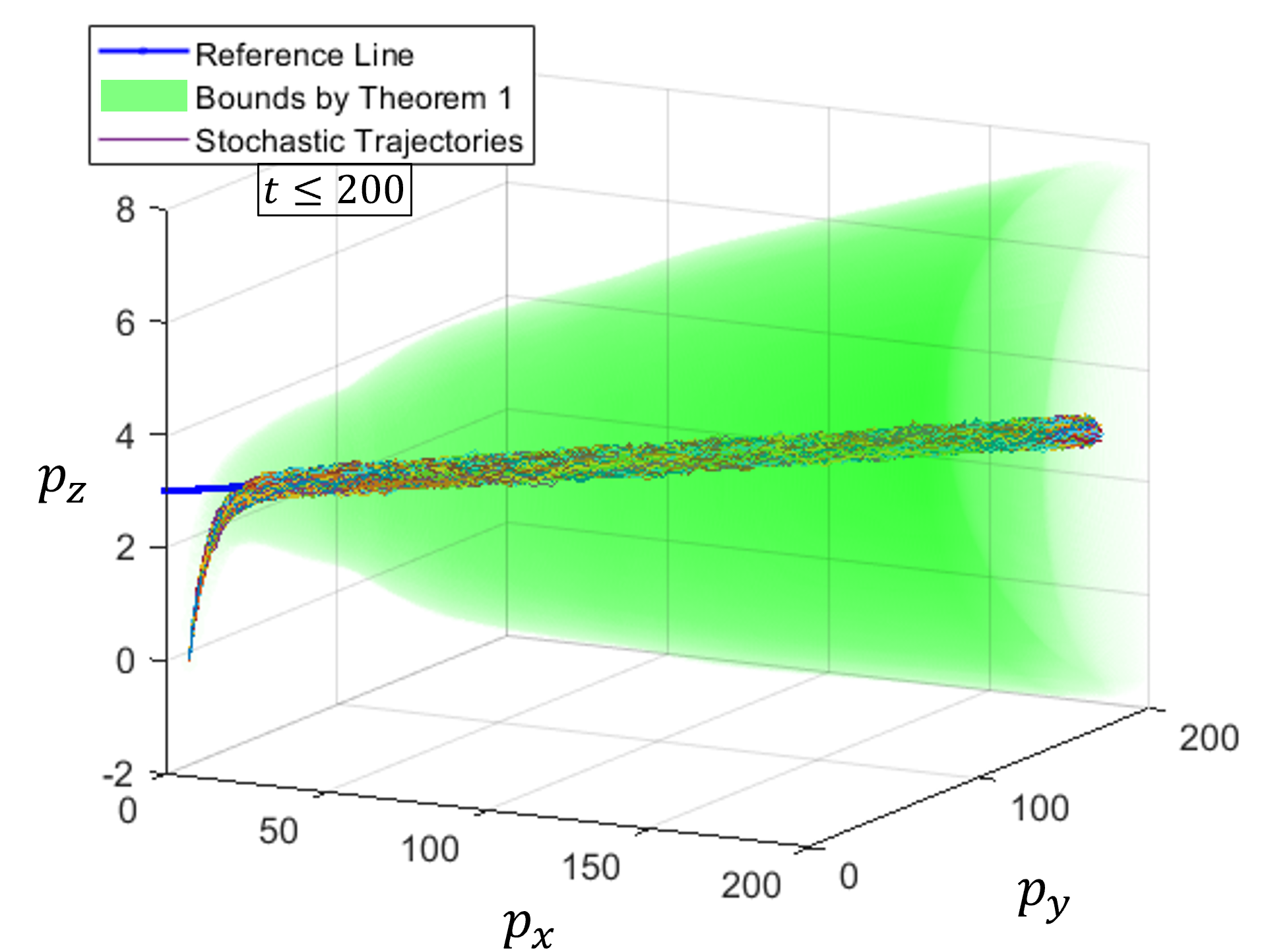}
  \caption{Path following task of the fixed-wing UAV. \textbf{First row:} two trajectories of the UAV \eqref{sys: UAV} at $t=50$ \textbf{(left)} and $t=150$ (right). The dashed blue line is the reference line and the dashed red curve is the deterministic trajectory. \textbf{Second row:} the top view \textbf{(left)} and the typical view \textbf{(right)} of the 2000 independent trajectories of the UAV \eqref{sys: UAV} and the $\delta$-PRS calculated by Theorem \ref{thm: reachable set} and Proposition \ref{prop: case 1}. Each trajectory contains 200 points. The solid blue line is the reference line and the green envelope is the calculated $\delta$-PRS.}
	\label{fig:UAV}
 \end{figure}

To verify our $\delta$-PRS, we generate a deterministic trajectory $x_t$ starting from $X_0$ on the associated deterministic system of \eqref{sys: UAV}, and then sample 2000 independent stochastic trajectories from the same $X_0$ during $t\leq200$. Two of the trajectories are visualized in the first row of Figure \ref{fig:UAV}. 

To make the probabilistic bounds tighter, we estimate a sequence of localized $L_t$ with respect to the $P$-weighted norm by simulated-based methods in \cite{9387079} with $P=diag(1,~1,~100,~50)$, and then compute the radius of the $\delta$-PRS by Theorem \ref{thm: reachable set} and the time-varying version of Proposition \ref{prop: case 1} with  $\|\cdot\|_{\mathbb{X}},\|\cdot\|_{\mathbb{U}}=\|\cdot\|_P$, $\varepsilon=1/16$ and $\delta=10^{-4}$. The union of the projections of the obtained $\delta$-PRS to the $p_x-p_y-p_z$ space for $t\leq200$ is visualized as the green envelopes in the second row of Figure \ref{fig:UAV}.
It is clear that all the sampled states are within the green envelopes, thus verifying our theoretical results.

\section{Conclusions}\label{sec: conclusion}
We present a unified and practical framework for computing the probabilistic reachable set (PRS) for discrete-time nonlinear stochastic systems, a generalization of our previous framework \cite{szy2024TAC} to the discrete-time system. 
%
A central idea of our approach is the separation strategy, which decouples the effect of deterministic inputs and stochastic noise on the PRS. The latter is captured by the difference between stochastic trajectories and their deterministic counterparts, termed stochastic deviation. 
%
A key technical innovation is a novel energy function called the Averaged Moment Generating Function, with which we establish a tight probabilistic bound on stochastic deviation. This bound is tight for general discrete-time nonlinear systems and is exact for linear systems. 
%
Based on the separation strategy, we can combine this tight bound with any existing methods for deterministic reachability to offer a PRS that is both flexible and effective with a high probability level. 
%
This PRS framework makes it possible to develop effective algorithms and products for control systems that require high probability guarantee on the PRS, such as safety-critical systems, and the AMGF leveraged in our work will open new research directions in multiple areas like control theory, finance, statistics, and machine learning.


\bibliographystyle{unsrt}        
\bibliography{main}

\appendix
\label{sec: appendix}

\section{Proof of Lemmas}

\subsection{Proof of Lemma \ref{lemma: AMGF_properties}} \label{app: lemma 4.1}
(1)-(2): See \cite{szy2024TAC}.

(3): By \eqref{eq: AMGF} we have
    \begin{equation}\label{eq: salf_x+w}
        \begin{split}
            \mathbb{E}_w\salf{x+w}&=\expectw{\ell\sim\mS^{n-1}}{\mathbb{E}_w e^{\lambda\innerp{\ell,x+w}}} \\
            &=\mbE_{{\ell\sim\mS^{n-1}}}\left(e^{\lambda\innerp{\ell,x}}\cdot \mbE_{w}e^{\lambda\innerp{\ell,w}}\right)
        \end{split}
    \end{equation}
    By the definition of sub-Gaussian vector, we know that 
    $$\mbE_{w}\left(e^{\lambda\innerp{\ell,w}}\right)\leq e^{\frac{\lambda^2\sigma^2}{2}}$$
    Therefore, \eqref{eq: salf_x+w} becomes
    $$ \mathbb{E}_w\salf{x+w}\leq e^{\frac{\lambda^2\sigma^2}{2}}\salf{x}$$
    which completes the proof.

\subsection{Proof of Lemma \ref{lemma: salf to sg}} \label{app: lemma 4.2}
 Define random vector $\tilde{X}=QX$, where $Q\sim\mathbb{U}^n$ is a random unitary matrix and $\mathbb{U}^n$ denotes the set of all the unitary matrices in $\R^{n\times n}$. By Lemma \ref{lemma: AMGF_properties}(i), we have that for any $\eta\in\mS^{n-1}$, 
        \begin{equation*}
        \begin{split}
            \salf{X}&=\expectw{\ell\sim\mS^{n-1}}{e^{\lambda\|X\|\innerp{\ell,\frac{X}{\|X\|}}}} =\expectw{\ell\sim\mS^{n-1}}{e^{\lambda\|X\|\innerp{\ell,\eta}}} \\
            &=\expectw{\ell\sim\mS^{n-1}}{e^{\lambda\innerp{\eta,\ell\|X\|}}} =\mbE_{Q\sim\mathbb{U}^n}\left(e^{\lambda\innerp{\eta,QX}}\right),
        \end{split}
        \end{equation*}
    where the last ``$=$'' uses the fact that $Q\ell\in\mS^{n-1}$ for any $\ell\in\mS^{n-1}$. Therefore, we obtain
    \begin{equation}\label{eq: Qx MGF half}
        \begin{split}
    &\mbE_{\tX}\left(e^{\lambda\innerp{\eta,\tX}}\right)=\mbE_X\mbE_Q\left(e^{\lambda\innerp{\eta,QX}}\right) \\
        = &\mbE_X\left(\salf{X}\right)\leq e^{\frac{\lambda^2\sigma^2}{2}},\quad\forall \lambda\in\R,~\forall \eta\in\mS^{n-1}.
        \end{split}
    \end{equation}
    Therefore, $\tilde{X}$ is sub-Gaussian with variance proxy $\sigma^2$. By Lemma \ref{lemma: concentration}, $\tilde{X}$ satisfies \eqref{eq: concentration norm}. 
    
    Finally, since $\|X\|=\|QX\|=\|\tilde{X}\|$ for any $Q\in\mathbb{U}^n$, we conclude that $X$ also satisfies \eqref{eq: concentration norm}. This completes the proof.

\end{document}